\documentclass{aa}  
\usepackage{graphicx}
\usepackage[varg]{txfonts}
\usepackage{natbib}
\usepackage{hyperref}

\newcommand\un[1]{{\,\rm #1}}
\newcommand\E[1]{\times10^{#1}}
\newcommand\rs[1]{_\mathrm{#1}}

\newcommand\g{$\gamma$}

\usepackage{xspace} 
\newcommand\tcb{T~CrB\xspace}

\usepackage{ulem} 
\usepackage[dvipsnames]{xcolor}

\begin{document}
	
\title{Probing the potential high-energy messengers\\ of 
the anticipated 
T~Coronae Borealis outburst}
\titlerunning{High-energy messengers from T Coronae Borealis}
\author{O. Petruk\inst{1,2}
    \and
    T. Kuzyo\inst{2}
	\and
	S. Orlando\inst{1}
    \and
    L. Chomiuk\inst{3}
	\and
    F. Bocchino\inst{1}
    \and
    M. Miceli\inst{1,4}
    \and
    S. Ustamujic\inst{1}
}

\institute{INAF - Osservatorio Astronomico di Palermo, Piazza del Parlamento 1, 90134 Palermo, Italy\\
    \email{oleh.petruk@inaf.it}	
	\and
	Institute for Applied Problems in Mechanics and Mathematics, National Academy of Sciences of Ukraine, Naukova St. 3-b, 79060 Lviv, Ukraine
    \and
    Center for Data Intensive and Time Domain Astronomy, Department of Physics and Astronomy, Michigan State University, East Lansing, MI 48824, USA
    \and
    Dipartimento di Fisica e Chimica E. Segrè, Università degli Studi di Palermo, Piazza del Parlamento 1, 90134, Palermo, Italy
}

\date{Received ...; accepted ...}

\abstract{T Coronae Borealis (T CrB) is a nearby recurrent nova expected to erupt in the near future, offering a unique opportunity to study particle acceleration and high-energy emission from novae in real time. We investigate the production of gamma-rays and neutrinos following the T CrB outburst by combining three-dimensional hydrodynamical simulations with a detailed diffusive shock acceleration model. 
Our simulations account for the complex circumbinary medium, including the red giant wind, equatorial density enhancement, and accretion disk. 
We compute spatially resolved spectra of accelerated protons and electrons at the forward shock, accounting for downstream velocity gradients and variations in shock properties. Using a multi-zone approach, we synthesize hadronic gamma-ray emission from proton-proton interactions, leptonic gamma-rays from inverse-Compton scattering, and the associated neutrino emission. 
We present predicted gamma-ray spectra, light curves, and images from our numerical models of T CrB, and assess their detectability with current gamma-ray and neutrino observatories. 
We find that the early high-energy emission is dominated by the ejecta, with the accretion disk significantly boosting the gamma-ray flux and particle normalization during the first hours after the outburst. 
By incorporating velocity gradients in the post-shock flow, we demonstrate that maximum particle energies can reach the PeV scale in high-energy explosion scenarios. 
We show that while GeV gamma-rays are prominent messengers, neutrino detection is feasible primarily in models with high explosion energy and high ambient density.}


\maketitle

\section{Introduction}
\label{tcorbor:sec-intro}

\tcb is a nearby recurrent nova that is likely to undergo a thermonuclear outburst in the very near future \citep[e.g.][]{2023MNRAS.524.3146S}. It is one of 10 known recurrent novae in our Galaxy \citep{2010ApJS..187..275S}.
It is a symbiotic binary, a close binary system in which a white dwarf (WD) accretes material from its red giant (RG) companion. Interestingly, this was the first nova for which the spectrum was taken \citep[][p.~44]{2014anst.book.....H}. Indeed, spectroscopy as a scientific tool was born in 1859, with analysis of the solar spectrum  \citep{1859BAM...oct..662K}. In 1866, the first spectroscopic observations of a nova were performed, and the target was \tcb \citep{1866MNRAS..26..275H}.

The outburst of \tcb is expected to occur around the year 2026, following the previous events of 1866, 1946 and possibly in 1787 and 1217 \citep{2023JHA....54..436S}. The observations of the nova could provide important data on the binary system, its environment, and the explosion itself. It will be one of the brightest novae of the century, primarily due to its very close distance
, which makes it an excellent target for multi-wavelength and multi-messenger campaigns.

Classical and recurrent novae are thermonuclear explosions that occur on the surfaces of white dwarfs that accrete material from its binary companions \citep{2021ARA&A..59..391C}. The sudden release of nuclear energy ejects material at velocities up to thousands km/s, creating strong shocks as the ejecta interacts with previously expelled material or with the ambient circumbinary medium (CBM). The shocks give rise to a broad range of electromagnetic emission, from radio to gamma rays, and potentially to high-energy neutrinos.

Radio observations have long been used to probe the ejecta structure, mass, and dynamics of novae \citep{2021ApJS..257...49C}. However, it is the discovery of GeV \g-ray emission by Fermi-LAT from nova from the symbiotic star V407~Cygni \citep{2010Sci...329..817A} and of TeV \g-rays from the recurrent nova RS Ophiuchi by MAGIC, H.E.S.S. and LST-1 \citep{2022NatAs...6..689A,2022Sci...376...77H,2025A&A...695A.152A} 
that significantly advanced our understanding of these systems. Subsequent detections of \g-rays from novae \citep[e.g.,][]{2025MNRAS.tmp.2143C}\footnote{According to the list of novae detected by Fermi telescope, there are currently 25 detections of Galactic novae with $>5\sigma$ significance and five with lower level. The list is compiled by Koji Mukai and is available online: 
\url{https://asd.gsfc.nasa.gov/Koji.Mukai/novae/latnovae.html}}
confirmed that novae are capable of accelerating particles to high energies. 

Gamma-ray observations of recent eruptions, such as those of classical novae V1369~Centauri, V5668~Sagittarii, V5856 Sagittarii, V906 Carinae \citep{2016ApJ...826..142C,2017NatAs...1..697L,2020NatAs...4..776A} and of the recurrent nova RS~Ophiuchi \citep{2022ApJ...935...44C}, have revealed complex, time-dependent \g-ray light curves and spectra that reflect the evolving shock, cosmic ray (CR) acceleration, and the density structure of the surrounding medium. 

Strong shocks are expected to produce high-energy neutrinos through proton–proton interactions in the CBM. Several searches for transient neutrino emission coincident with novae have been performed using IceCube data 
\citep{2023ApJ...953..160A,2023arXiv230715372T}. Although no significant detections have yet been reported, theoretical studies suggest that modern and the next generation of instruments could detect neutrinos from nearby novae, particularly from recurrent systems embedded in dense environments \citep{2023JCAP...03..015G,2025PhRvD.111d3035W}. 
To achieve this goal, IceCube Collaboration is preparing for observations of \tcb \citep{2025arXiv250707096T}.

From a theoretical point of view, semi-analytical modeling and detailed numerical simulations explore the nova outburst \citep[e.g.,][]{2025A&A...698A.251J,2025ApJ...982...89S,2025ApJ...991..200W} as well as the interplay between nova ejecta, shock dynamics, and particle acceleration \citep[e.g.,][]{2015MNRAS.450.2739M,2021ApJ...910..134G,2025A&A...704A.144O}. 

An exploration of various models of \tcb offers a potential framework for the analysis of future observations. For example, a set of different numerical models of nova explosions developed by \citet{2025A&A...698A.251J} 
aimed to be a diagnostic tool to characterize the thermonuclear runaway, the explosion dynamics, and the resulting ejecta of a future \tcb nova event. 
A set of three-dimensional (3D) hydrodynamical (HD) numerical models of the \tcb binary system presented by \citet[][hereafter Paper I]{2025A&A...704A.144O} focuses on a nova outburst in a complex environment and the subsequent evolution of the nova remnant. In that work, the thermal X-ray emission produced during the remnant’s expansion is synthesized, providing detailed insights into the range of diagnostic information that observations with current X-ray facilities can yield. 

In the present paper, we aim to explore potential high-energy messengers from \tcb following the outburst. These include \g-rays and neutrinos produced through proton–proton interactions, as well as leptonic \g-ray emission arising from inverse-Compton (IC) scattering. 
We focus on the phase that follows the thermonuclear explosion, when the forward shock propagates outwards and interacts with the CBM. 
Our approach is built upon the 3D HD models of \tcb developed in our Paper I, incorporating a description of relativistic particle spectra and synthesizing the production of high-energy \g-rays and neutrinos.

\citet{2025arXiv251222338S} recently provided predictions for the \tcb high-energy emission using semi-analytical and 1D frameworks; our work extends these by utilizing full 3D HD simulations to capture the impact of environment geometry and properties of CR acceleration.

\section{Our approach}
\label{tcorbor:sec-method}

\subsection{Numerical hydrodynamical models}

In Paper I, we developed several detailed 3D HD models of the recurrent nova \tcb. As determined in previous studies \citep[e.g.][]{2025ApJ...983...76H}, the binary consists of a WD with mass $1.37M_\sun$ and a RG companion with mass $0.69M_\sun$ and radius $64R_\sun$; the binary separation is $0.9\un{a.u.}$. 
The circumbinary medium around \tcb in the models is composed of three main components: the RG wind with an equatorial density enhancement (EDE) and the accretion disk, having characteristics consistent with recent observations of \tcb. The WD outburst produces a strong shock expanding into this complex environment. The 3D HD numerical models are nonstationary and therefore help to investigate the propagation of the blast wave and the associated X-ray emission. These simulations enable detailed predictions of X-ray diagnostics, like light curves, spectra, and emission maps, with instruments currently in orbit, including XMM-Newton/RGS and XRISM/Resolve. 

In this paper, we examine the generation of cosmic rays and the accompanying \g-rays and neutrinos as well as their detectability, during the shock's propagation through the CBM after the outburst using a few models from the Paper I. 
In these numerical models, several input parameters are fixed to values determined in the literature (e.g., WD and RG masses, orbital parameters). Others (ejected mass and energy, characteristics of CBM components) define a parameter space that was narrowed down using radio data (Paper I, Sect.~2). 

For the purposes of the present paper, we selected four models (Table~\ref{tcb:table-HDmodels}) out of the 14 considered in Paper I. 
The models differ mainly by the density distributions in the CBM, which affect the outcome of hadronic interactions. 
A spherical wind and EDE has a hydrogen number density prescribed as
\begin{equation}
 n\rs{rg}(\mathbf{r})=n\rs{w}\left(\frac{1\un{pc}}{r}\right)^2+
 n\rs{ede}\exp\left[-\left(\frac{x}{h_x}\right)^2-\left(\frac{y}{h_y}\right)^2-\left(\frac{z}{h_z}\right)^2\right]
\end{equation}
where $\mathbf{r}=(x,y,z)$ is a position vector with the absolute value of a distance from the RG center $r$, $n\rs{w}$ is the wind density at a distance $1\un{pc}$, $n\rs{ede}$ is the peak density of EDE, $h_x$, $h_y$, $h_z$ are length-scales for the EDE in the respective directions.
The accretion disk is centered on the WD with the shape described as 
$h\rs{disk}(\varrho)=h\rs{max}\left(\varrho/R\rs{disk}\right)^2$  \citep{2000ApJ...534L.189H}
where $h\rs{disk}$ is the height from the $xy$-plane, $\varrho$ is the distance in this plane, the maximum radius $R\rs{disk}=0.5\un{a.u.}$ and the maximum vertical extent $h\rs{max}=0.3\un{a.u.}$ Its density is modeled as
$ n\rs{disk}(\mathbf{r})=10^3\, n\rs{rg}(\mathbf{r})$.
All models in Table~\ref{tcb:table-HDmodels} assume an ejected mass $3\E{-7}M_\sun$, $n\rs{w} = 10^{-3}\un{cm^{-3}}$, and $h_y=h_x$. 

\begin{table}
\caption{Parameters in HD models of \tcb. The reference model is RUN04. 
The last column indicates the presence of the accretion disk in the numerical setup.}
\begin{tabular}{cccccc}
\hline
Model & $E\rs{bw}$ & $n\rs{ede}$ & $h_z$ & $h_x$ & disk \\
    & $10^{43}\un{erg}$ & $10^7\un{cm^{-3}}$ & $10^{13}\un{cm}$ & $10^{13}\un{cm}$ & \\
\hline
RUN04 & 3.0 & 0.1 & 4.5 & 4.0 & Y \\
RUN03 & 3.0 & 10. & 4.5 & 8.0 & Y \\
RUN10 & 10. & 1.0 & 3.0 & 8.0 & Y \\
RUN14 & 3.0 & 0.1 & 4.5 & 4.0 & N \\
\hline
\end{tabular}
\label{tcb:table-HDmodels}
\end{table}

The 3D Cartesian computational grid is centered at the WD location, with the WD explosion energy $E\rs{bw}$.
In all models, the initial radius of the blast wave is $14R_\sun$, which corresponds to about 30 minutes or 0.02 days after the explosion. The ejecta within this radius has a uniform temperature and density, with some random fluctuations in $\rho$ added on top of it; the flow velocity is distributed as $v(r)\propto r$. 
The initial post-shock velocity of the ejecta material in the models RUN04, RUN03, RUN14 is $4100\un{km/s}$ that is in the range $4000-4700\un{km/s}$ of the highest velocities derived from the spectra of \tcb in 1946 around its maximum \citep{1946PASP...58..159M,1946PASP...58..196H}. 
In the model RUN10, this velocity is $7500\un{km/s}$. 

Numerical simulations were performed with the PLUTO code \citep{2007ApJS..170..228M}. 
Further details on the exploration of the parameter space, the simulation setup, and numerical methods, together with relevant illustrations, are presented in Paper I. 
We synthesize the observables by summing up emissivities over grid cells. The resulting spectra, images, and light curves depend on the 3D structure of the nova remnant and thus vary across different models.

\subsection{Maximum momentum of accelerated particles}
\label{tcb:sect-pmax}

We consider the diffusive shock acceleration of CRs at the forward shock. 
Since our numerical models are HD, we add a magnetic field (MF) during post-processing.
The strength of the amplified MF $\delta B_1$ is given by the saturated level of the \citet{2004MNRAS.353..550B} non-resonant instability with the magnetic energy
\begin{equation}
 \frac{\delta B_1^2}{4\pi}\simeq\xi\rs{cr}\rho_1V^3/c
 \label{tcorbor:eq-dB}
\end{equation}
where $\xi\rs{cr}$ is a fraction 
of the kinetic energy of the shock transferred to CRs, $\rho_1$ is the pre-shock density and $V$ is the shock speed. Subscripts 1 and 2 denote the upstream and downstream regions, respectively.
The orientation of $\delta B_1$ is assumed to be random and, therefore, MF is compressed at the shock by a factor $\sigma\rs{\delta B}=\sqrt{11}$. 
MF of this strength does not affect the HD structure, as its energy density amounts to only $\xi\rs{cr}V/c\sim 0.05$ of the kinetic energy density and is therefore dynamically unimportant.

The maximum momentum $p\rs{max}$ of accelerated particles is defined by the minimum of values given by time-limited acceleration $p\rs{m1}$, time-limited growth of the fastest-growing \citet{2004MNRAS.353..550B} modes $p\rs{m2}$, and radiative losses $p\rs{m3}$. 

We consider the Bohm diffusion coefficient for particle acceleration %
\begin{equation}
 D(p)=\frac{wcp}{3\,q\,\delta B}
\end{equation}
where $w$ is the particle speed. In the upstream and downstream, this formula is applied using subscripts 1 and 2, respectively.

By considering the equation for the acceleration time $t\rs{acc}=3\beta\rs{a}D_1/V^2$ \citep{1983RPPh...46..973D} and equating it to the amount of time since eruption $t\rs{acc}=t$, we have 
\begin{equation}
 p\rs{m1}=\frac{q\,\delta B_1V^2t}{\beta\rs{a}c^2},
 \label{tcorbor:eq-pm1}   
\end{equation}
where $\beta\rs{a}$ accounts for the time particles spend upstream and downstream. We consider the shock compression $\sigma=4$, the compression of the turbulent MF $\sigma\rs{\delta B}$, and Bohm diffusion; therefore $\beta\rs{a}=2.94$. 

The time-to-grow for the Bell instability results in the following limitation for the maximum momentum of protons \citep{2015APh....69....1C,2020APh...12302492C}:
\begin{equation}
	p\rs{m2}\simeq\frac{\xi\rs{cr}q\pi^{1/2}}{5\Lambda} 
	\frac{4}{4-\omega}
	R \rho_1^{1/2}\left(\frac{V}{c}\right)^2,
 \label{tcorbor:eq-pm2}   
\end{equation}
valid for any diffusion type, where $\rho_1\propto r^{-\omega}$ is the pre-shock density with $\omega=2$, $\Lambda\approx 9$ for $p\rs{max}\sim 10\un{TeV}$. 

The maximum momentum due to losses is given by the equation $t\rs{los}=t\rs{acc}$ where the energy-loss time is $t\rs{los}=E/\dot E$ with the energy-loss rate $\dot E$. The synchrotron loss time is $t\rs{syn}\simeq\left[A\rs{e}B^2p\right]^{-1}$ where $A\rs{e}=4.7\E{7}\un{cgs}$. Then, for electrons,  
\begin{equation}
 p\rs{m3}=\left[\frac{qV^2\delta B_1}{A\rs{e}\beta\rs{a}c^2B^2}\right]^{1/2}
 = \left[\frac{qV^2}{A\rs{e}c^2\beta\rs{b}\delta B_1}\right]^{1/2}
 \label{tcorbor:eq-pm3e} 
\end{equation}
where $B^2\simeq \delta B_1^2 \zeta_1+\delta B_2^2 \zeta_2$ where $\zeta_1=t\rs{acc1}/t\rs{acc}$ and $\zeta_2=t\rs{acc2}/t\rs{acc}$ are fractions of time the particles spend upstream and downstream. This results in 
$\beta\rs{b}=\sigma(\sigma+1)/(\sigma-1)=6.67$. 

Protons suffer losses due to hadronic interactions, with the loss time 
$t\rs{pp}\simeq \left[A\rs{p}n\rs{H}\right]^{-1}\approx 7\E{7}/n\rs{H}\un{yr}$ where $A\rs{p}=0.5c\sigma\rs{pp}$ and $n\rs{H}$ is the number density of target protons. The cross-section $\sigma\rs{pp}$ depends very weakly on the proton momentum \citep{2006PhRvD..74c4018K,2014PhRvD..90l3014K}. We take $\sigma\rs{pp}\simeq 30\un{mbarn}$ and therefore $t\rs{pp}$ is independent of $p$.
Radiative losses of protons could be effective if $n\rs{H}\gtrsim 7\E{7}/t\rs{yr}\,\un{cm^{-3}}$. 
The shock could probe such high densities during the first days after the outburst (figure~3 in Paper I). 
Then, by equating $t\rs{pp}=t\rs{acc}$, for protons,
\begin{equation}
 p\rs{m3}=\frac{qV^2\delta B_1}{A\rs{p}\beta\rs{a} c^2n\rs{H}}.
\label{tcorbor:eq-pm3p}   
\end{equation}

\subsection{Spectrum of accelerated particles}
\label{tcb:sect-accCR}

The classic theory of diffusive shock acceleration predicts a power-law distribution of momenta, while the interpretation of observed spectra of non-thermal emission often requires particle distributions that deviate from a pure power law (e.g., a broken power law or a log parabola). \citet{2024A&A...688A.108P} and \citet{2025arXiv251018763C} demonstrated that the flow velocity profile spatially varying downstream on the length scale involved in particle acceleration affects the momentum distribution of accelerated particles.

\citet{2024A&A...688A.108P} derived an analytic solution for the particle spectrum under such conditions. \citet{2025arXiv251021988P} applied the solution to the post-adiabatic shock when the radiative losses of plasma trigger essential changes in the structure of the flow. The spatial distribution of the flow velocity may also deviate from the uniform one in the early stages after the nova or supernova explosions \citep{1982ApJ...258..790C,1998ApJ...497..807D}. Therefore, in the present paper, we take this effect into account.

Let us consider the case where the flow velocity $u$ in the reference frame of the shock has a constant gradient $du/dx$ after the jump at the shock:
\begin{equation}
 u(x)=\left\{
 \begin{array}{ll}
  u_1>0& \mathrm{if\ } x<0\\
  u_2=u_1/\sigma& \mathrm{if\ } x=0\\
  u_2+[du/dx]x& \mathrm{if\ } x>0
 \end{array}
 \right.
 \label{tcorbor:eq-u2}
\end{equation}
where $u_1$ and $u_2$ are the uniform pre-shock and the immediate post-shock values of the flow velocity, the shock compression factor is $\sigma=4$, the shock is located at the origin $x=0$, and $x>0$ corresponds to the downstream region. 
We do not consider the back-reaction of CRs on the flow. We assume a spatially constant diffusion coefficient $D\propto p$ and the diffusion distance $x\rs{p}\simeq D/u_2$. 
The momentum distribution of accelerated particles in such a model is given by an
analytic solution \citep[equation (9) in][]{2025arXiv251021988P}. It differs from a common shape by a factor that involves the gradient of the flow speed $[du/dx]$ and the diffusion distance $x\rs{p}$. We use this factor in the present paper, that is, the momentum distribution of protons is assumed to be 
\begin{equation}
 N(p)=K p^{-s}\exp\left(-\left[\frac{p}{p\rs{max}}\right]^\beta\right)
 \times \left[1-\frac{\mu \bar x\rs{p}(p)}{\sigma-1}\right]^{\,\delta}
 \label{tcb:eqN}
\end{equation}
where $\mu\equiv [d\bar u/d\bar x]$, $\bar x=x/R$, $\bar u=u/u_2$, $s=(\sigma+2)/(\sigma-1)$, $\delta={[(3\kappa-1)\,\sigma+1]/(\sigma-1)}$ and we take $\beta=1$. Note that $\mu$ is dimensionless by construction.

The flow speed gradient is $\mu\approx 1$ for a Sedov shock \citep{1950RSPSA.201..159T}. 
Depending on the sign of the velocity gradient, the spectrum becomes softer for $\mu>0$ \citep[Sedov shock, figure~3 in][]{2024A&A...688A.108P} or harder for $\mu<0$ \citep[the partially-radiative shocks,][]{2025arXiv251021988P}.
In very young shocks, $\mu$ could be larger or smaller than zero (see, e.g., figure~1 in \citet{1982ApJ...258..790C}, figure~3 in \citet{1998ApJ...497..807D} or animation A1 in \citet{2021MNRAS.505..755P}). In the present paper, we calculate $\mu$ from the numerical data (see Appendix~\ref{tcorbor:app-mu} for details).

A typical approach, which assumes a power-law momentum distribution with an exponential cut-off, is reproduced by setting $\mu=0$. 
To see the effect of the gradient of $u(x)$, we consider two cases: the one that accounts for the local gradients of $u$ in Eq.~(\ref{tcb:eqN}) and the second, a more common one, with $\mu=0$. In the rest of the paper, we will mark the first case as $\mu\neq 0$.

After the injection momentum $p\rs{in}$, the CR spectrum decreases with $p$. Therefore, we calculate the normalization $K$ as%
\begin{equation}
 \xi\rs{in} n\simeq \int_{p\rs{in}}^{p\rs{max}}Kp^{-s}dp
\end{equation}
by assuming $p\rs{in}\ll p\rs{max}$ and $s=2$ in this formula. This results in
\begin{equation}
	K\simeq \xi\rs{in} n_2 p\rs{in}
    =\xi\rs{in} n_2\, \chi \sqrt{2m\rs{H}k\rs{B}T\rs{2p}}
 \label{tcb:defK}   
\end{equation}	
where $\xi\rs{in}$ is the injection efficiency, i.e., a fraction of the post-shock particles that are involved in the acceleration process, $p\rs{in}$ 
is the injection momentum, which is $\chi=3.7$ times \citep[e.g.][]{2005MNRAS.361..907B} the thermal momentum for protons. The temperature of the proton population $T\rs{2p}\approx T_2$, where $T_2\propto V^2$ is the post-shock mean plasma temperature from Rankine–Hugoniot conditions, which depends on the location. We keep $\xi\rs{in}$ constant across the shock surface at a given $t$. The value of $\xi\rs{in}$ is taken to provide the overall acceleration efficiency $\hat \xi\rs{cr}=0.1$ (see the details in Appendix~\ref{tcorbor:app-xicr}). 

The momentum distribution of electrons has the same shape as the proton spectrum except for $p\rs{max}$ and the normalization $K$, which is $K\rs{ep}$ times that of the proton spectrum normalization. 
In addition, in the cells where the maximum momentum of electrons is determined by the radiative losses (i.e., by $p\rs{m3}$), one should take $\beta=2$ and multiply (\ref{tcb:eqN}) by the term 
\begin{equation}
 \left[1+0.523\left(\frac{p}{p\rs{max}}\right)^{9/4}\right]^2
  \label{tcb:factor-losses}   
\end{equation}
which slightly modifies the electron distribution around $p\rs{max}$ \citep{2007A&A...465..695Z}. 

The normalization of the electron momentum distribution is given by an expression similar to Eq.~(\ref{tcb:defK}) with the change $\sqrt{m\rs{H}T\rs{2p}}\rightarrow (\mu\rs{e}/\mu\rs{H})\sqrt{m\rs{e}T\rs{2e}}$. The value of $K\rs{ep}$ is therefore $K\rs{ep}\simeq (\mu\rs{e}/\mu\rs{H})(m\rs{e}/m\rs{p})^{1/2}(T\rs{2e}/T\rs{2p})^{1/2}$, i.e., could be in the range $\simeq0.01\div0.001$. Since the downstream electron temperature $T\rs{2e}$ is not an output of our simulations, we treat $K\rs{ep}$ as a parameter with the value $K\rs{ep}=0.01$ unless specified otherwise.

\subsection{Propagating particles}
\label{tcb:sect-escCR}

The root-mean-square (rms) propagation distance of escaping particles is given by
$\ell\equiv\sqrt{\langle \Delta r^2\rangle} = \sqrt{2\, D\, \Delta t}$
where $\Delta t$ is the time after escape and $\Delta r$ is the distance from the shock. 
To estimate $\ell_1$ outside of a remnant, we consider $D_1=2\E{26}\,(pB\rs{ref}/B_1 p\rs{ref})^{0.5}\un{cm^{-2}s^{-1}}$ with $p\rs{ref}=1\un{GeV}$ and the MF strength $B_1\simeq B\rs{ref}=3\un{\mu G}$. 
In one month, particles with energy $10\un{TeV}$ could propagate $\ell_1\simeq 0.1\un{pc}$ out of the remnant. 
If we take $D_2\simeq D_1\cdot (B\rs{1}/B\rs{2})^{0.5}$ with $B_2\simeq 100\un{\mu G}$ inside a remnant, then the rms distance $\ell_2\simeq0.04\un{pc}$, in one month. 
The radius of the remnant in our simulations is about $400\un{a.u.}\approx 0.002\un{pc}$ after one year of expansion. Thus, escaping protons could illuminate the entire interior through $pp$-collisions as well as the ambient medium at large distances around the remnant. 

The propagation time for distances comparable to the radius of the remnant is considerably shorter than its age. 
That is, the CR diffusion dominates downstream advection. 
Therefore, the evolution of the proton spectrum inside the remnant and outside in its vicinity is given by the solution $g(p,r)$ of a stationary diffusion equation $D{\partial^2 g(r)}/{\partial r^2}=0$ with the boundary conditions $\partial g/\partial r=0$ at $\pm\infty$ and $g(p)=g\rs{o}(p)$ at the shock, that is, $g(p,r)=g\rs{o}(p)$. 
In other words, we assume that the CR spectrum generated at the shock at a given time $g\rs{o}(p)\equiv N(p)$ propagates 'immediately` and without changes over the length scales comparable with the remnant's size. 

\subsection{Synthesis of observables}
\label{tcorbor:sect-observables}

Emission synthesis is performed using a multi-zone approach by calculating emissivities in each grid zone of a numerical model and then summing them up over the emitting volume. 
We adopt the distance $d=890\un{pc}$ to \tcb \citep{2021AJ....161..147B} and an orbital inclination $55\degr$ \citep{2025ApJ...983...76H}. 

The \g-ray emissivity in the nova interior (that consists of ejecta and shocked CBM) is calculated using the Naima package \citep{2015ICRC...34..922Z} with the Monte Carlo model \verb+Pythia8+ for hadronic interactions.
The IC \g-rays are calculated on the black-body radiation fields of the Cosmic Microwave Background (temperature $2.72\un{K}$, energy density $0.261\un{eV/cm^{3}}$) and of the RG wind (temperature $\sim 3000\un{K}$, energy density $1\un{eV/cm^{3}}$ corresponding to the distance of $\sim30$ radii of RG). 
Possible effects from the nova radiation field, namely, generation of \g-rays from IC scatterings on this field as well as the plasma opacity for GeV-PeV photons are considered in Sect.~\ref{tcorbor:sec-nova-radiation}.

Other leptonic processes, such as synchrotron emission and non-thermal bremsstrahlung, may contribute to the hard X-ray/low-energy \g-ray emission of novae. They are expected to be subdominant in GeV-TeV photon generation and are therefore not considered in the present paper. We also neglect emission from secondary electrons (a product of proton-proton interactions).

The WD is not resolved in our simulations and is therefore excluded from the synthesis of observables. We also exclude the RG star, since the hydrogen in the RG star is neutral, while the production of hadronic emission requires free protons. The shock is rapidly damped after entering the very dense material, so it cannot ionize the RG star. 
Indeed, the average number density of the RG in our model is $n\rs{star}=3\E{18}\un{cm^{-3}}$ while the wind density around it is about $n\rs{wrg}=5\E{8}\un{cm^{-3}}$. Therefore, the shock speed in the RG becomes $(n\rs{wrg}/n\rs{star})^{1/2}\sim 10^{-5}$ times its pre-entering value $\sim 1000\un{km/s}$, i.e., almost zero. The sound speed in the CBM is about $15\un{km/s}$ and increases with temperature in the red giant envelope, resulting in the shock Mach number rapidly falling below unity.


CRs escaping upstream into the CBM and then into the interstellar medium could produce \g-ray emission by interacting with the ambient matter or photons. As we will see later, the contribution to the light curve from the shocked CBM is negligible during the period of \tcb detectability. Therefore, we do not consider the high-energy emission from the ambient medium. In any case, constructing a dedicated CBM-ISM model would be required to perform a proper synthesis of such an emission. 

The total spectrum of emitted neutrinos is a sum of spectra of muonic and electronic neutrinos and antineutrinos.  
It follows quite closely the \g-ray spectrum produced in the same $pp$ collisions but with a bit lower high-energy cutoff (Appendix~\ref{tcorbor:app-1}). Therefore, we use the hadronic \g-ray spectrum calculated with the Naima package with the neutrino maximum energy $E\rs{\nu, max}\approx0.8E\rs{\gamma, max}$ as a reasonable proxy for the emitted neutrino spectrum. The oscillations on the way to the observer change the composition of the neutrinos. However, the spectrum of all-flavor neutrinos arrived to the Earth is the same as the total spectrum emitted at the source.

\begin{figure*}
  \centering 
  \includegraphics[width=0.97\textwidth]{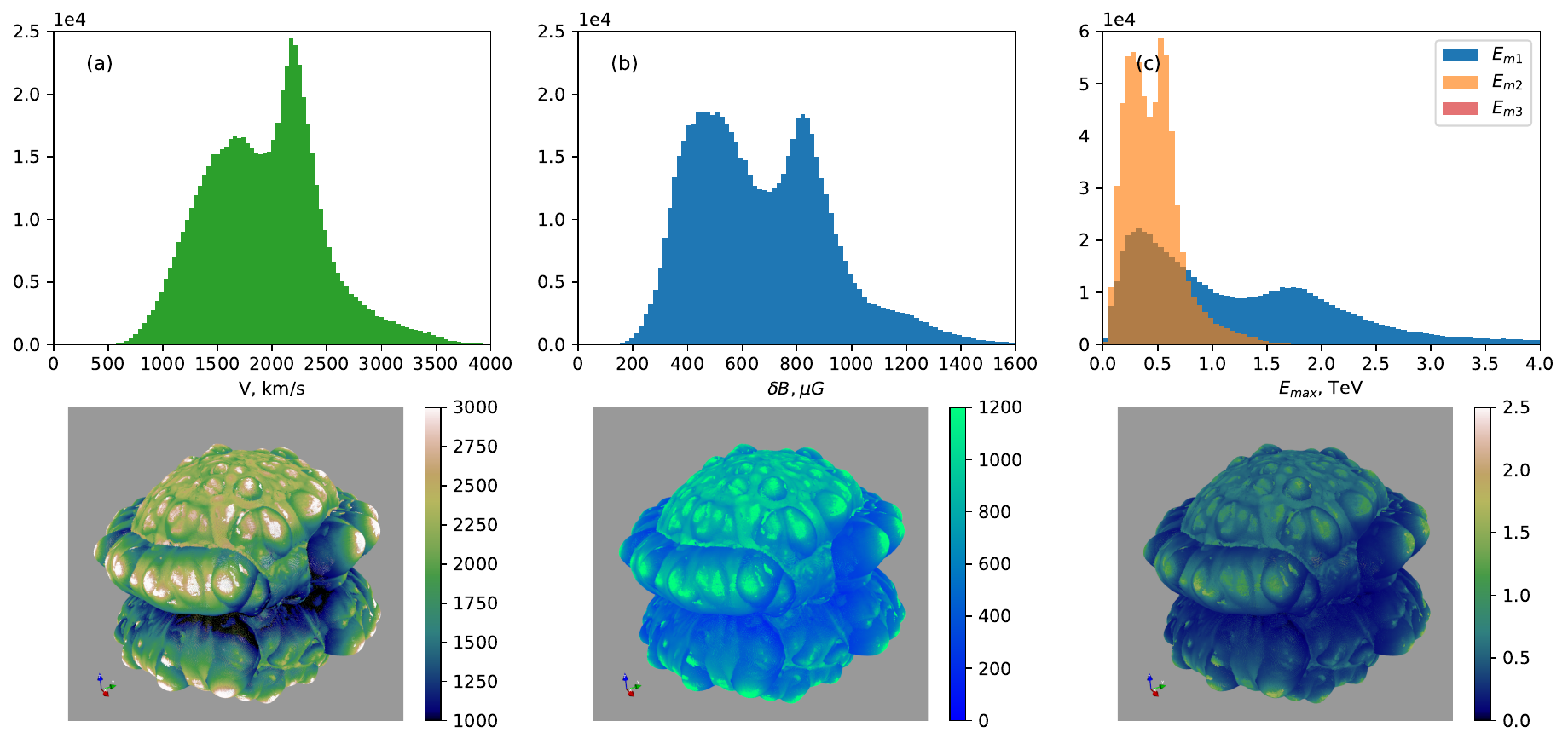}\\
  \includegraphics[width=0.97\textwidth]{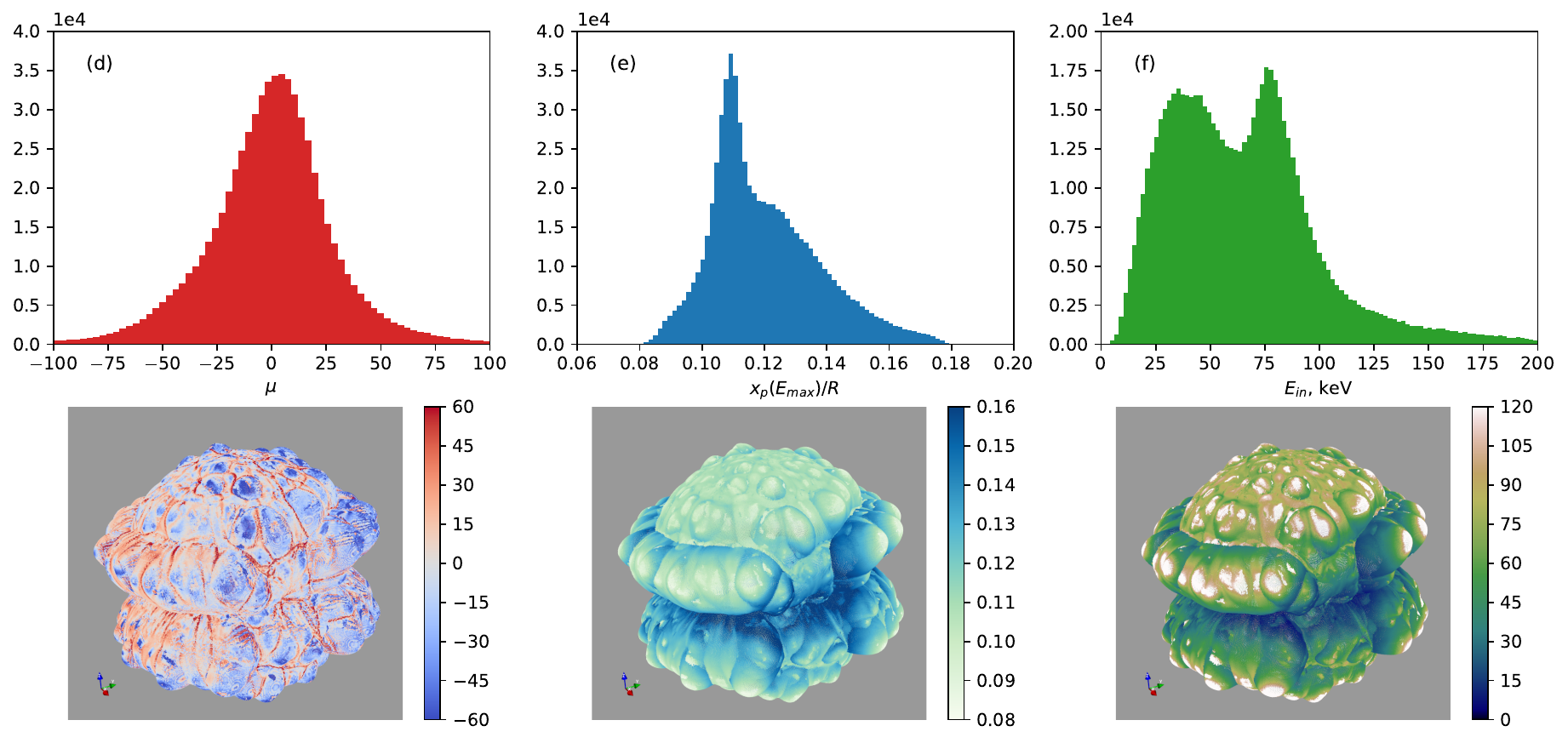}
  \caption{%
       Histograms (rows 1 and 3) and corresponding spatial distributions (rows 2 and 4) for parameters determining the shape of the proton spectrum, across the shock surface for RUN04, $t = 153$ days, $\hat \xi\rs{cr}=0.1$, $\mu\neq 0$. The vertical axes on histograms show the number of pixels. The color scales for the surface plots have the same units as the horizontal axes in histograms.
       Top two rows: the shock speed $V$ (left), the disordered magnetic field $\delta B$ (center), the maximum energy of protons $E\rs{m1}$ and $E\rs{m2}$ (right). $E\rs{m3}$ is not shown here because it is a few orders of magnitude higher than $E\rs{m1}$ and $E\rs{m2}$. The surface distribution shows $\min(E\rs{m1},E\rs{m2})$.
       Bottom two rows: the gradient of the flow speed $\mu$ (left), the diffusion distance $x\rs{p}(p\rs{max})/R$ for protons with momentum $p\rs{max}$ (center), the injection energy $E\rs{in}$ (right). 
       The distribution of the flow gradient is characterized by $\mu = 0.2 \pm 27.9$. The blue line on the histogram for $\mu$ represents a Gaussian with the mean $0.2$ and standard deviation $27.9$. 
       The shock radius varies, being smaller in the orbital plane and larger out of it, but the spread in shock radius across all possible directions is rather small:  
       $R = 252 \pm 25\un{a.u.}$.
  }
  \label{tcorbor:fig-hist-r-V-B}
\end{figure*}

\begin{figure}
  \centering 
  \includegraphics[width=\columnwidth]{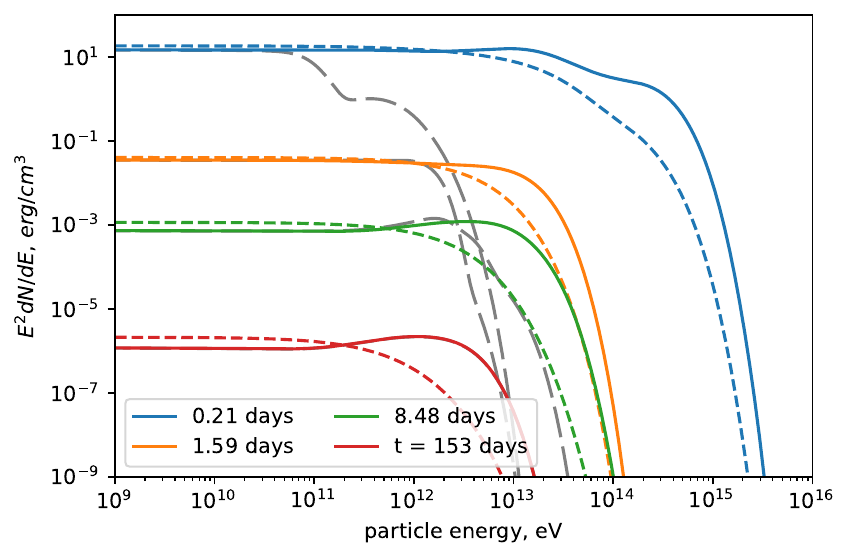} 
  \caption{%
       The average proton energy distributions for several time moments in the RUN04 model.  
       The solid lines correspond to the model that accounts for variations in $\mu$ across the shock surface, and the short-dashed lines are for $\mu = 0$. The gray long-dashed lines show the electron spectra in the case of variable $\mu$ and $K\rs{ep}=1$, enabling a direct comparison with protons. 
       The dashed and solid lines are not aligned at low energies because the spectra have the same acceleration efficiency $\hat \xi\rs{cr}=0.1$. At lower energies, all these spectra are $E^{-2}$ power laws, in agreement with Eq.~(\ref{tcb:eqN}).
  }
  \label{tcorbor:fig-proton-specta}
\end{figure}

\section{Results}
\label{tcorbor:sec-res}

\subsection{Closer look at the particle acceleration}
\label{tcorbor:sec-acc-cons}

The distribution of parameters characterizing the momentum distribution of protons is shown in Fig.~\ref{tcorbor:fig-hist-r-V-B} for a snapshot in RUN04, which is our reference model of \tcb. 
Most parameters appear to have a bimodal distribution, reflecting locations at the forward shock with high (in the directions out of the equatorial plane) and low shock speeds (in the equatorial plane that is denser), as shown by the 3D representations in Fig.~\ref{tcorbor:fig-hist-r-V-B} a. 

In particular, the histogram of amplified MF strength at the day $153$ (Fig.~\ref{tcorbor:fig-hist-r-V-B} b) shows two clear peaks, around $\delta B\simeq 0.5$ and $0.8\un{mG}$, with some cells having values above $1\un{mG}$. 
At the earlier times, MF is significantly higher. 
The distributions of $E\rs{m1}$ and $E\rs{m2}$ values (from Eqs.~\ref{tcorbor:eq-pm1} and \ref{tcorbor:eq-pm2}) are also bimodal (Fig.~\ref{tcorbor:fig-hist-r-V-B} c). Our analysis shows that the maximum energy of protons is limited by the time of MF growth, i.e. by $p\rs{m2}$ values over most of the shock surface during most of the simulation time. 

The variation in the shock radius across different directions is visible from the shape of the surface distributions.  
On average, the spread of $R$ is rather small: $R = (1.22 \pm 0.12)\cdot10^{-3}\un{pc}$. It is asymmetric with respect to the average: the fraction of the remnant's surface with shorter radii (in the vicinity of the equator) is small compared to the surface with larger $R$, where interaction of the shock with EDE is not efficient.

The maximum distance from the shock the particles may probe while being accelerated $x\rs{p}(p\rs{max})$ is about 10\% of the shock radius (Fig.~\ref{tcorbor:fig-hist-r-V-B} e). The flow speed certainly is not constant on such length scales and, therefore, the CR spectrum should be affected by the non-uniformity of the flow velocity, that is, by $\mu$, which becomes a necessary component for modeling the particle momentum distribution. 

In Fig.~\ref{tcorbor:fig-hist-r-V-B} d, the distribution of $\mu$ has a mean value $0.2$ and a standard deviation $27.9$. 
A clear correlation can be seen from the surface distribution of $\mu$: cells with negative $\mu$ typically correspond to large $V$, while positive $\mu$ is mostly found where $V$ is small. 
The variation of $\mu$ in the numerical model arises from the structure of the flow. 
Indeed, there are turbulent cells (`bubbles') and $\mu$ is positive around their edges and negative at their `caps'.
As mentioned above, the negative $\mu$ leads to a harder particle spectrum, while the positive $\mu$ causes a cutoff at momenta lower than $p\rs{max}$ given by formulae in Sect.~\ref{tcb:sect-pmax}. 

For this snapshot of the RUN04 model, the injection energy, which determines the normalization of the particle spectrum through corresponding $p\rs{in}$ in Eq.~(\ref{tcb:defK}), varies by roughly a factor of four across the shock due to variations in density and shock speed (Fig.~\ref{tcorbor:fig-hist-r-V-B} f). 

The momentum distribution of protons, averaged over the shock surface, reflects the spatial variations of all these parameters across the remnant's surface. 
It is shown in Fig.~\ref{tcorbor:fig-proton-specta} for a few time moments. 
It is calculated for each cell on the shock surface using Eq.~(\ref{tcb:eqN}), considering the local parameters, and then averaged. 
One can see that taking into account the flow speed gradient $\mu$ could lead to larger attainable maximum energies of CRs than in the case of a uniform downstream flow with $\mu=0$ (cf. solid and short-dashed lines in Fig.~\ref{tcorbor:fig-proton-specta}). For example, the value of $E\rs{max}=\min(E\rs{m1},E\rs{m2},E\rs{m3})\approx 0.2\un{TeV}$ at $t=153\un{days}$ for the average spectrum (Fig.~\ref{tcorbor:fig-proton-specta}, red short-dashed line), while the actual particle spectrum has effectively larger $E\rs{max,eff}\approx 2\un{TeV}$ (Fig.~\ref{tcorbor:fig-proton-specta}, red solid lines). This is an order of magnitude difference. The situation is similar at other times as well.

\begin{figure}
  \centering 
  \includegraphics[width=\columnwidth]{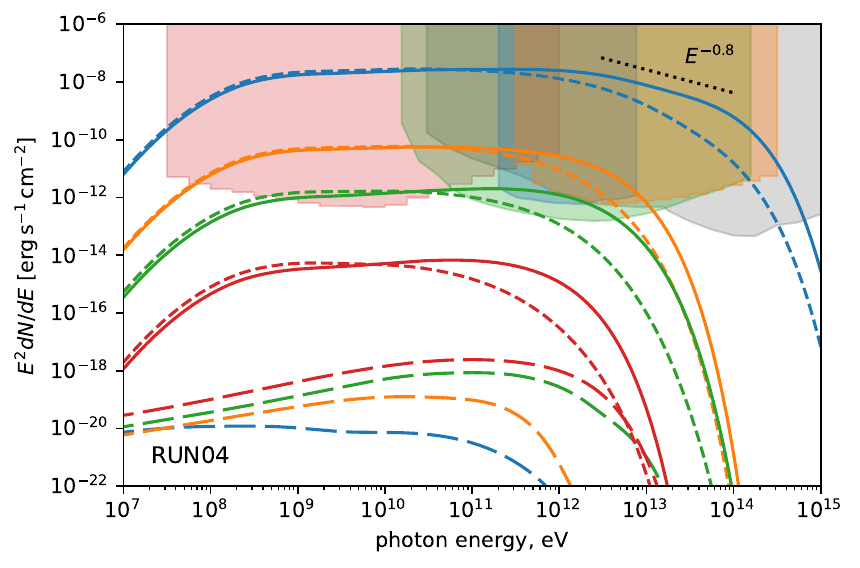} 
  \caption{%
       The \g-ray spectra for the same time moments as in Fig.~\ref{tcorbor:fig-proton-specta}.  
       The solid and short-dashed lines are for hadronic \g-rays. 
       The solid lines correspond to the model that accounts for variations in $\mu$ across the shock surface, and the short-dashed lines are for $\mu = 0$. The long-dashed lines represent the IC \g-ray spectra for the case $\mu\neq 0$ and $K\rs{ep}=0.01$. Spectra of neutrinos are similar to hadronic \g-ray spectra with the high-energy cutoff at about 0.8 times smaller energy.
       The shaded regions correspond to the differential sensitivity of Fermi LAT (pink, 10 yrs, taken from \href{https://www.slac.stanford.edu/exp/glast/groups/canda/lat_Performance.htm}{LAT Performance}), H.E.S.S. (blue, 50 hours, \citet{2015arXiv150902902H}), LHAASO (gray, 1 yr, \citet{2019arXiv190502773C}), ASTRI (red, 50 hours, \citet{2022icrc.confE.884L}), CTAO North Alpha (green, 50 hours, \citet{2022icrc.confE.884L}). 
       The sensitivity of VERITAS is quite similar to that of H.E.S.S. 
  }
  \label{tcorbor:fig-pp-specta}
\end{figure}

As the average kinetic energy density and CBM density decrease, both the maximum cosmic-ray energy and the spectral amplitude decline with time (Fig.~\ref{tcorbor:fig-proton-specta}).
The spectral shape changes due to the multi-zone nature of the system and the temporal and spatial variability of its parameters.
It is interesting that during the first few hours the maximum energy attainable by protons could be close to the PeV scale (Fig.~\ref{tcorbor:fig-proton-specta}, solid blue line). 
Such a high $E\rs{max}$ is reached in our models of \tcb despite the explosion energy, which is considerably lower than in supernovae. It should be a highly amplified magnetic field and a relatively large density that cause such cosmic-ray energies in novae. 

Fig.~\ref{tcorbor:fig-proton-specta} also demonstrates the energy distribution of electrons (gray lines). We see that the difference between the maximum energies of electrons and protons decreases with time. They differ significantly at early times due to the strong radiative losses experienced by electrons in the highly amplified MF. For electrons, radiative losses predominantly limit $p\rs{max}$ during the first few days. Subsequently, the two values $p\rs{m2}$ and $p\rs{m3}$ compete on the shock surface, and after a few months, $p\rs{m2}$ provides the lowest value of the maximum momentum for electrons almost everywhere.  

\subsection{Gamma-rays from the basic model}

Let us now look at the \g-ray emission from the reference model RUN04. We stress that the spectra and light curves obtained from our 3D HD models of \tcb are fundamentally multi-zone in nature. That is, the total emission from a model is an integral of emissivities with spatially variable parameters.

The spectra are shown in Fig.~\ref{tcorbor:fig-pp-specta}. 
It is apparent that hadronic \g-ray radiation (solid and short-dashed lines) is dominant, while the contribution from the IC process (long-dashed lines) is negligible. With time, the difference between the two components decreases, likely due to decreasing density and magnetic field, which enhance the leptonic emission and reduce the hadronic one. 

Similarly to the proton energy distribution, the \g-ray spectrum extends to higher photon energies when downstream flow-speed non-uniformity is taken into account (solid versus short-dashed lines in Fig.~\ref{tcorbor:fig-pp-specta}). 

Interestingly, the shape of the \g-ray spectrum may have some features due to its multi-zone nature. In particular, the solid blue line in Fig.~\ref{tcorbor:fig-pp-specta} has a spectral break around a few TeV with the change of the $dN/dE$ slope from $E_\gamma^{-2}$ to $E_\gamma^{-2.8}$. 

The emission spectra are compared with the sensitivities of several experiments to provide an approximate estimate of the nova's detectability. The location of \tcb in the northern sky provides VERITAS, ASTRI, CTAO North, LHAASO with favorable conditions for \g-ray observations, as \tcb can appear nearly at the zenith.

\begin{figure}
  \centering 
  \includegraphics[width=\columnwidth]{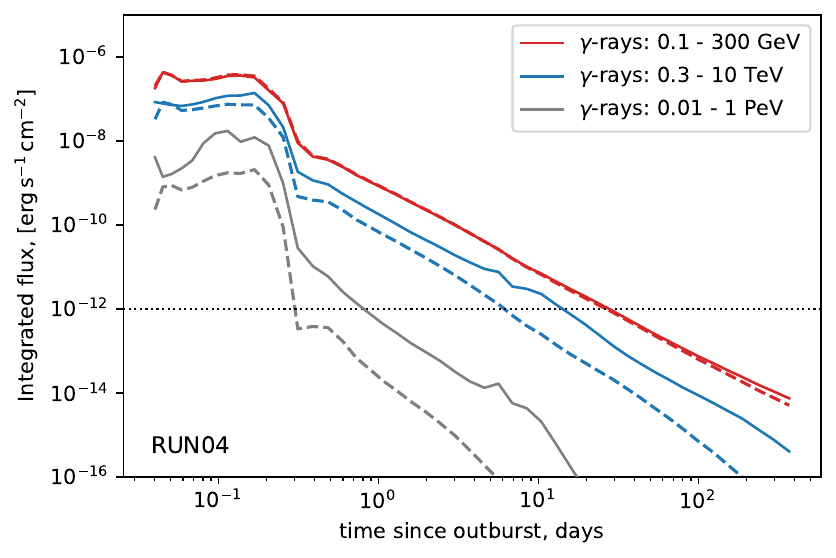} 
  \caption{%
       The light curves for \g-rays in three photon energy ranges estimated from the RUN04 model of \tcb.  
       The solid lines correspond to the model that accounts for variations in $\mu$ across the shock surface, and dashed lines correspond to the model with $\mu = 0$. The thin horizontal line marks the sensitivity limit of different instruments (see  Fig.~\ref{tcorbor:fig-pp-specta}). Its intersection with the light curve provides a rough estimate of a nova's visibility duration. 
  }
  \label{tcorbor:fig-pp-gamma-ray-light-curve}
\end{figure}

\begin{figure}
  \centering 
  \includegraphics[width=\columnwidth]{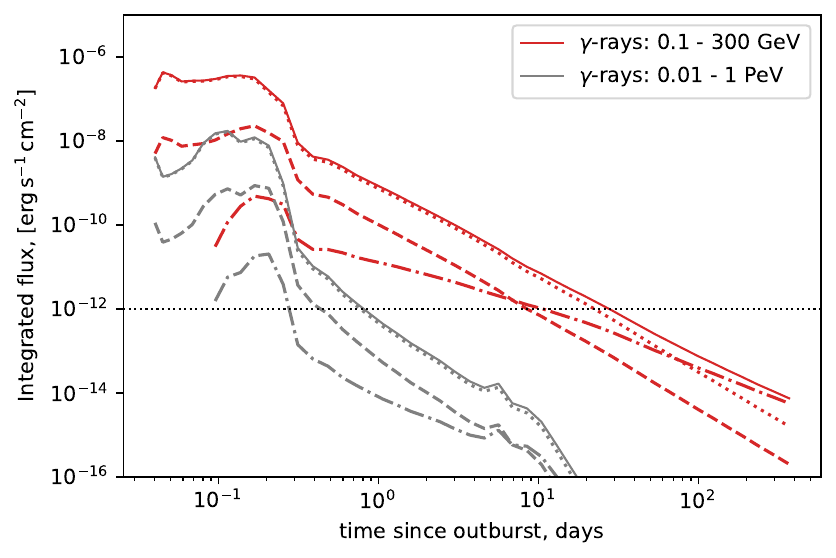} 
  \caption{%
       Contribution of different components to the \g-ray light curves in two photon energy ranges for the RUN04 model of \tcb with $\mu\neq 0$. Contributions shown are from the shocked ejecta (dotted line), the shocked disk material (dashed line), and the shocked CBM plasma (dot-dashed line). Each solid line is the sum of these components; they are the same as solid lines in Fig.~\ref{tcorbor:fig-pp-gamma-ray-light-curve}. 
  }
  \label{tcorbor:fig-pp-gamma-ray-lc-contributions}
\end{figure}

Light curves for \g-rays produced in the RUN04 model are shown in Fig.~\ref{tcorbor:fig-pp-gamma-ray-light-curve} for a few photon energy ranges: $0.1-300\un{GeV}$ (corresponds to Fermi LAT), $0.3-10\un{TeV}$ (H.E.S.S., VERITAS, MAGIC), $0.01-1\un{PeV}$ (LHAASO). 
Solid and dashed lines compare the two approaches, $\mu\neq 0$ and $\mu=0$. Differences are noticeable in the hard bands that include the high-energy cutoff: the flux is higher when the flow speed gradient is taken into account. This is in agreement with the \g-ray spectra, Fig.~\ref{tcorbor:fig-pp-specta}. All light curves demonstrate rather sharp decay around 8 hours (0.3 days) when the shock has just run out of the accretion disk. This is related to the influence of the disk on the CR acceleration: the maximum energy $p\rs{max}\propto \rho_1^{1/2}$. 
Indeed, the dominant contribution to the \g-ray flux comes from the ejecta material highlighted by the relativistic protons (dotted lines in Fig.~\ref{tcorbor:fig-pp-gamma-ray-lc-contributions}). The direct contribution from the proton-illuminated disk material (dashed lines) is one-two orders of magnitude lower than from the ejecta. The fraction of \g-rays generated in the shocked ambient CBM is also negligible (dot-dashed lines). It rises, however, over time to the level where it becomes dominant after about three months in the Fermi range but then the nova has dropped below detectability. This behavior of components contributing to light curves in the RUN04 model is similar to the soft X-ray light curve (Figure~7 in our Paper I). However, in hard X-rays, the shocked disk is dominant during the first few hours (the same figure). 

\begin{figure}
  \centering 
  \includegraphics[width=0.56\columnwidth]{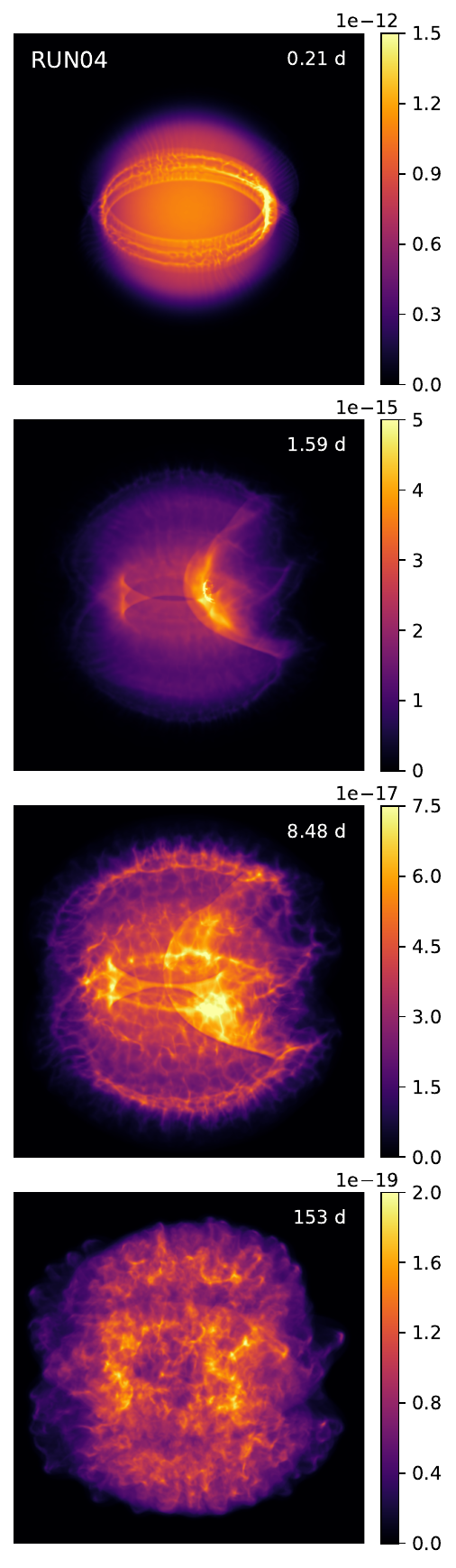}
  \caption{%
       The hadronic \g-ray images of \tcb corresponding to cases shown by the solid lines in Fig.~\ref{tcorbor:fig-pp-specta}, in photons with energies $0.1-300\un{GeV}$. Time is shown by white digits on the plots. 
       The data-cube with emissivities is rotated first by the $20\degr$ angle about the vertical $z$-axis (this corresponds to an arbitrary choice of an orbital phase of the binary system at the time of the outburst) and then on the orbital inclination angle $55\degr$ about the horizontal $x$-axis (to match the inclination of the system versus us). 
       The $y$-axis is directed out of the observer. 
       The color scales are in units $\un{erg\, s^{-1}\,cm^{-2}}$. 
  }
  \label{tcorbor:fig-pp-images}
\end{figure}

Accelerated CRs also illuminate the unshocked ambient medium, by escaping upstream and producing \g-rays there. However, this contribution is expected to be comparable with the shocked CBM's contribution, that is, marginal compared to the ejecta, at least during the period of about one month when \tcb can be detected. For this reason, we do not consider \g-ray emission from the unshocked CBM in the present paper.

Fig.~\ref{tcorbor:fig-pp-images} shows the spatial distribution of \g-rays in the RUN04 model.
The approach used to derive the images is similar to that employed in the synthesis of hadronic \g-ray images of supernova remnants \citep{2012MNRAS.419.1421B}. Namely, the \g-rays arise from interactions of relativistic protons with the non-relativistic protons within the remnant. The images are shown at full resolution to highlight the regions where most of the emission originates.

The first image in Fig.~\ref{tcorbor:fig-pp-images} corresponds to the time when a fraction of the shock is still within the accretion disk and has just reached the surface of the RG; the right part of the image is a bit brighter due to the higher density of the RG wind around the star. The second image shows the situation shortly after the shock completely overruns the RG star (which occurred on day 1). The third plot represents the system close to the time when it is about to fall below the detectability threshold. The fourth image shows a remnant after 5 months; it is characterized by rather diffuse emission, with only weak traces of the structures present earlier.

\begin{figure}
  \centering 
  \includegraphics[width=0.849\columnwidth]{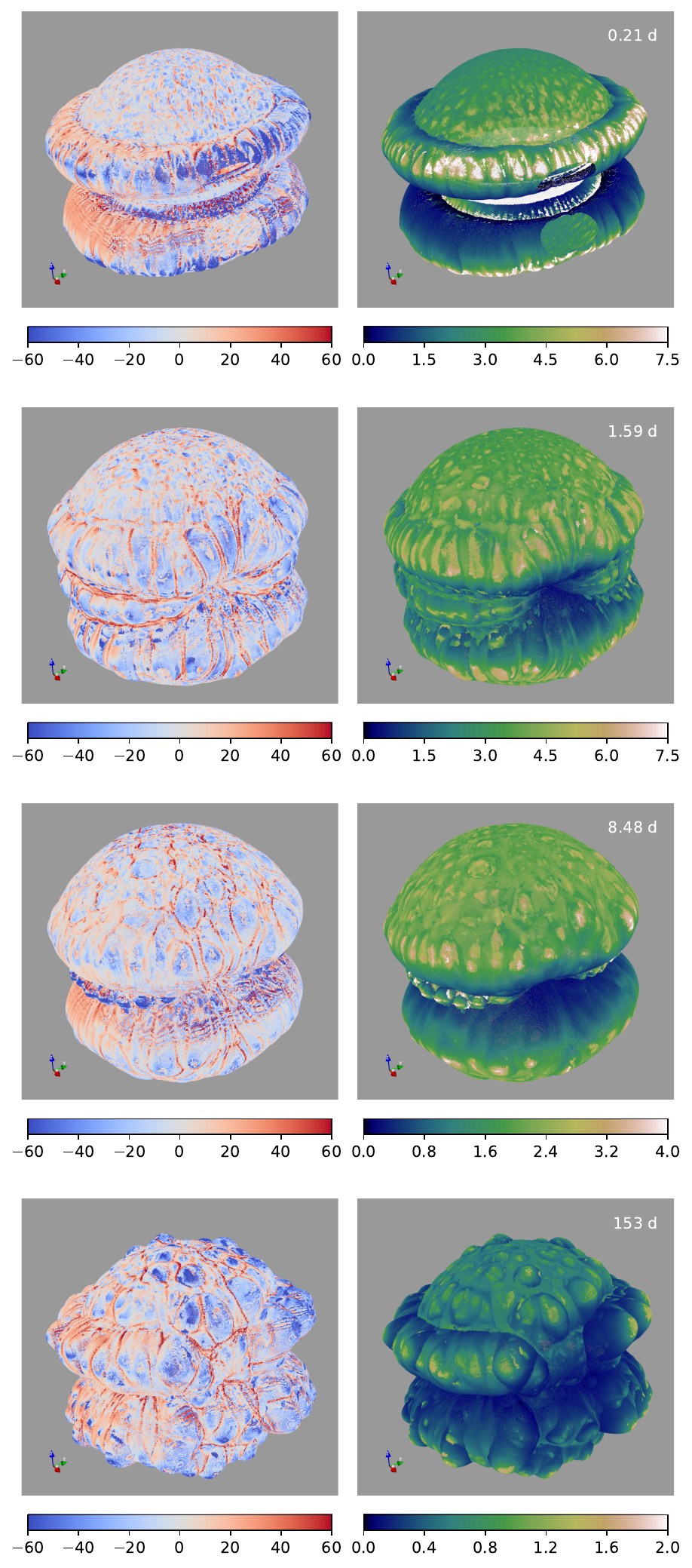}
  \caption{%
       The distribution of $\mu$ (on the left) and $E\rs{max}$ in TeV calculated with formulae from Sect.~\ref{tcb:sect-pmax} (on the right) across the shock surface for the same HD model and times as in Fig.~\ref{tcorbor:fig-pp-images}. 
  }
  \label{tcorbor:fig-3d-mu-emax}
\end{figure}

Fig.~\ref{tcorbor:fig-3d-mu-emax} confronts the spatial distributions of the flow properties with the maximum energy of CRs, which form the emission map. Note that the values shown are $E\rs{max}$ given by formulae in Sect.~\ref{tcb:sect-pmax}. The actual highest energies attainable by protons could be larger (if $\mu$ is negative) or smaller (for positive $\mu$) than $E\rs{max}$.  
Namely, Fig.~\ref{tcorbor:fig-3d-mu-emax} shows variations of $E\rs{max}$ and $\mu$ across the shock surface for the same moments of time as in Fig.~\ref{tcorbor:fig-pp-images}. In particular, the first snapshot clearly shows that the part of the shock propagating in the accretion disk results in the largest values of $E\rs{max}$ for protons, due to the very high density there.   
In other regions, there is a correlation between regions with negative $\mu$ (bluish colors in the left plots) with the highest values of $E\rs{max}$ (white and yellow colors in the right plots). 
Protons can reach the highest energies if the two factors happen at the same place: the high values of $E\rs{max}$ (calculated without taking into account non-uniformity of the downstream flow) and the negative $\mu$, which further increase $E\rs{max}$. 

The physical reason for such behavior may be easier to understand from the last plot at $t=153\un{days}$. 
Indeed, the maximum values of $E\rs{max}$ are found around caps of bubbles. The largest absolute values of the negative $\mu$ (blue color) are also located there. In contrast, bubbles are surrounded by narrow `valleys' with large positive $\mu$ (red color) and $p\rs{max}$ reaches its lowest values there. Detailed consideration of the internal structure of these bubbles shows clearly that they correspond to the ejecta protrusions, where the plasma from deeper downstream moves faster than the plasma near the forward shock. These are the locations where the gradient of the flow speed $\mu$ is negative. The protrusions are developed due to instabilities at the contact discontinuity between ejecta and shocked ambient medium.
This clearly demonstrates that the effect from the non-uniformity of the plasma flow downstream (it is represented by $\mu$ in our calculations) is quite important for CR acceleration and is physically relevant for the early phases of a remnant evolution. 


\begin{figure*}
  \centering 
  \includegraphics[width=0.54\columnwidth]{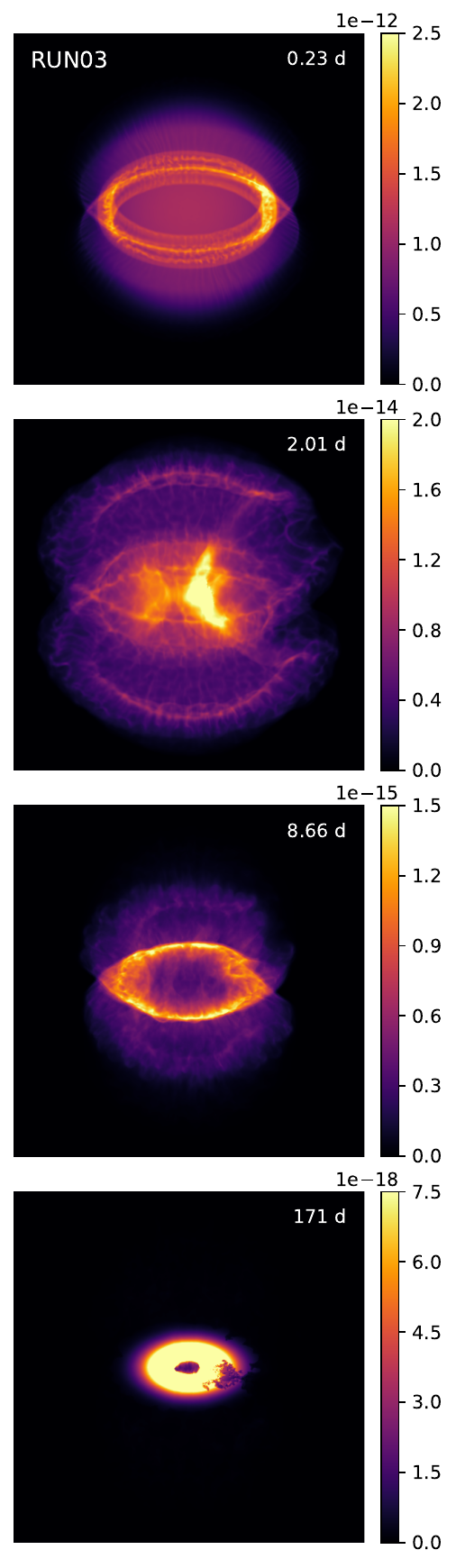}\ 
  \includegraphics[width=0.54\columnwidth]{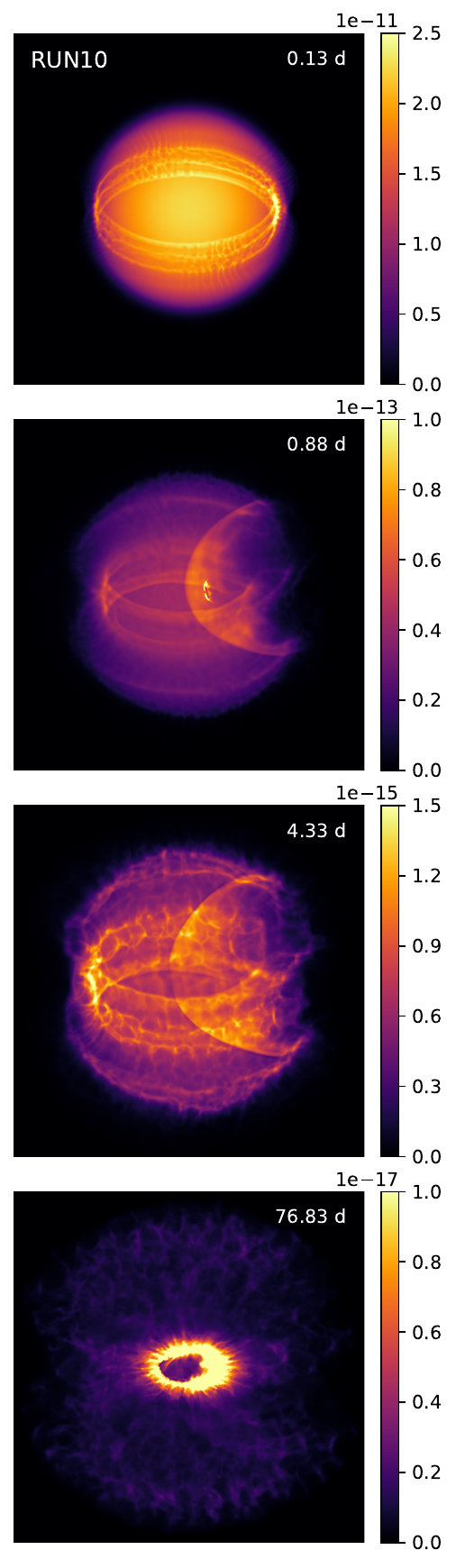}\ 
  \includegraphics[width=0.54\columnwidth]{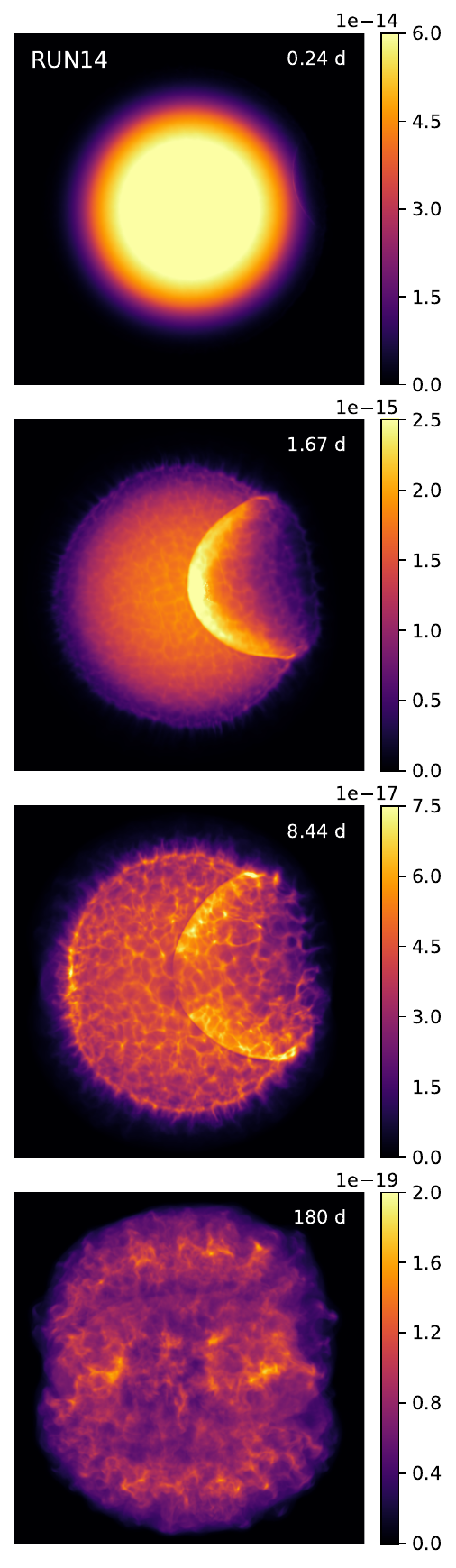}
  \caption{%
       Hadronic \g-ray images for models RUN03 (left column), RUN10 (center), RUN14 (right). 
       The photon energy range, system orientation, and units on the color scale are the same as in Fig.~\ref{tcorbor:fig-pp-images}. The white number on each plot is the time since the outburst. 
       Physical sizes of the black squares are the same within a row and correspond to the sizes in Fig.~\ref{tcorbor:fig-pp-images}. The extent of the nova remnants may therefore be directly compared in a given row. 
      }
  \label{tcorbor:fig-pp-images-models}
\end{figure*}
\begin{figure}
  \centering 
  \includegraphics[width=\columnwidth]{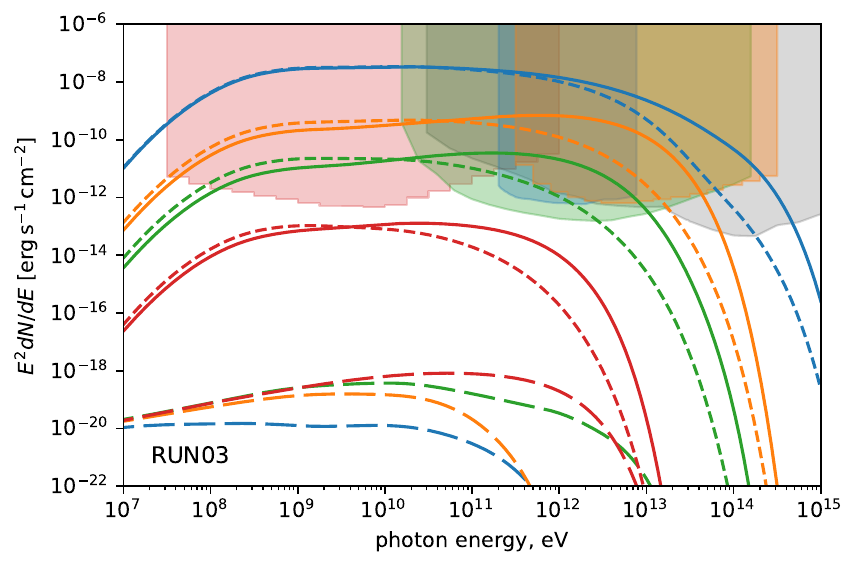} 
  \includegraphics[width=\columnwidth]{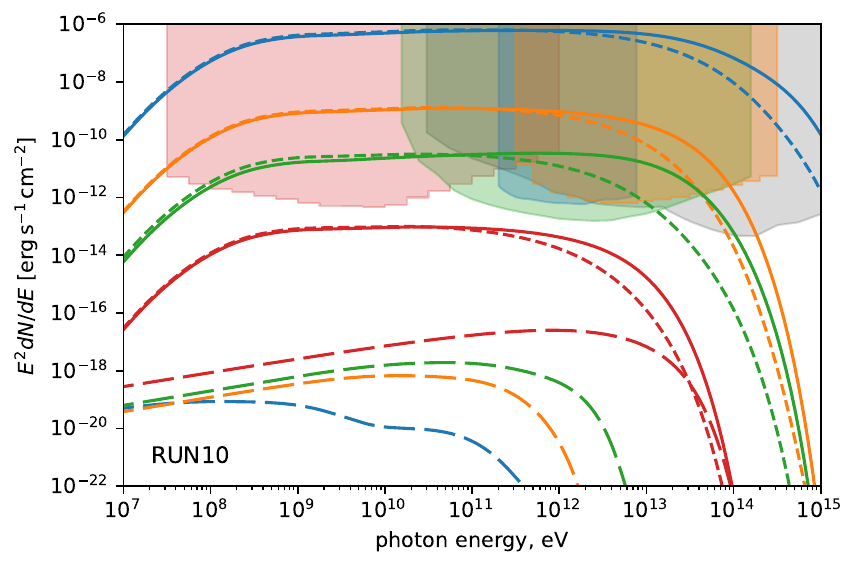} 
  \includegraphics[width=\columnwidth]{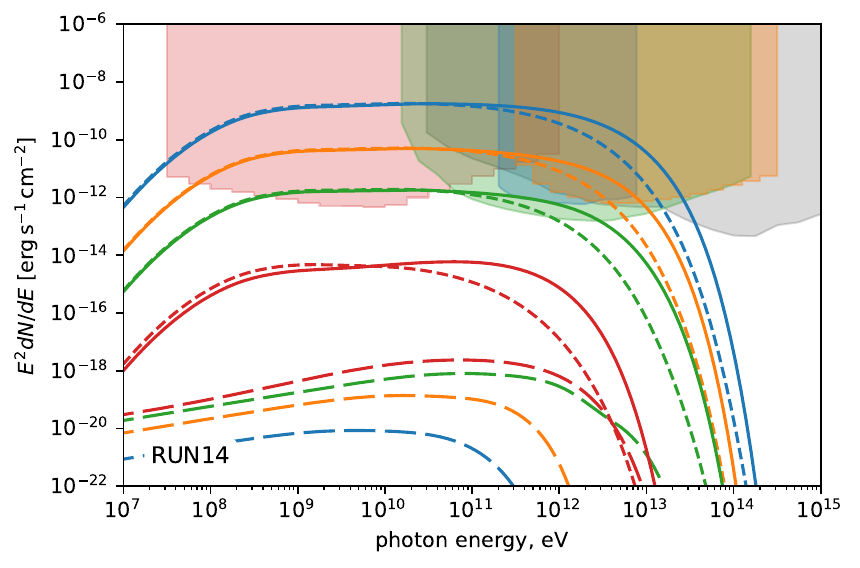} 
  \caption{%
       Gamma-ray spectra for models of \tcb RUN03, RUN10, RUN14 (top, middle, bottom) for the time moments shown in Fig.~\ref{tcorbor:fig-pp-images-models}. Lines have the same meaning as in Fig.~\ref{tcorbor:fig-pp-specta}.
  }
  \label{tcorbor:fig-pp-gamma-ray-spectra-models}
\end{figure}
\begin{figure}
  \centering 
  \includegraphics[width=0.99\columnwidth]{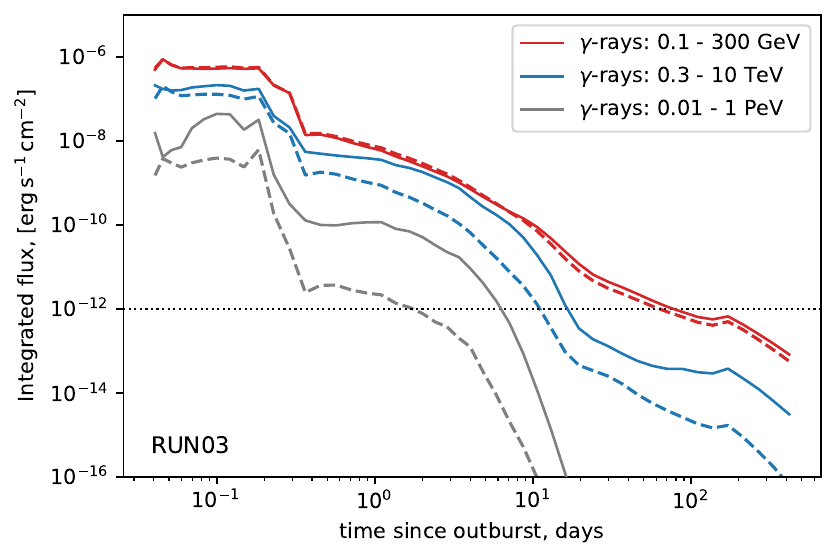} 
  \includegraphics[width=0.99\columnwidth]{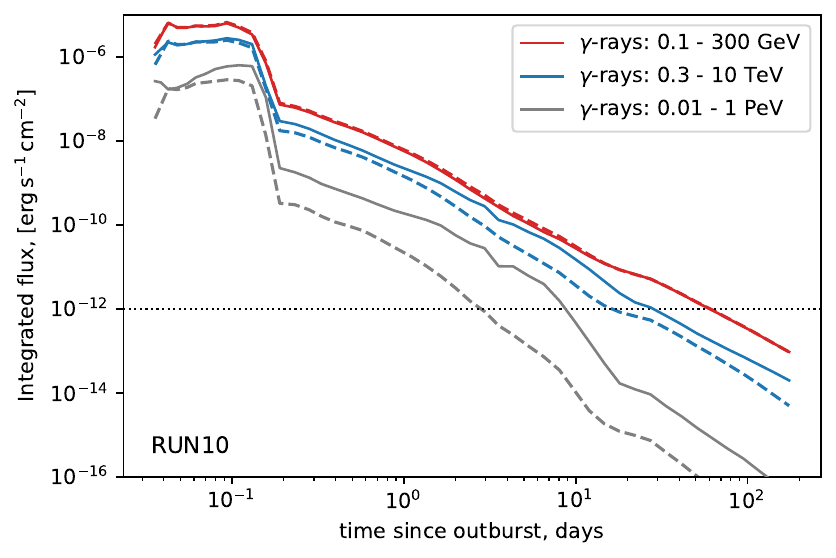} 
  \includegraphics[width=0.99\columnwidth]{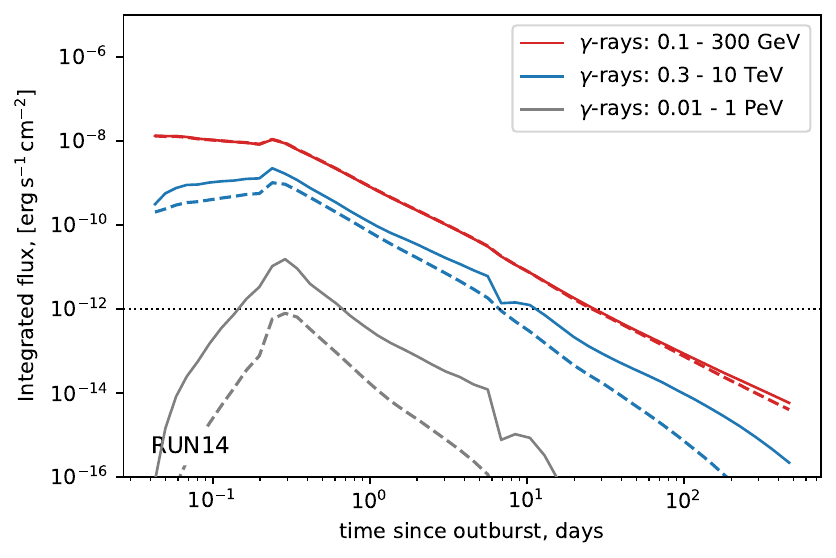} 
  \caption{%
       Gamma-ray light curves for models RUN03 (top), RUN10 (middle), and RUN14 (bottom). Lines have the same meaning as in Fig.~\ref{tcorbor:fig-pp-gamma-ray-light-curve}.
  }
  \label{tcorbor:fig-pp-gamma-ray-lc-models}
\end{figure}

\subsection{Gamma-rays from other models}

Let us now compare the ability of other models of \tcb to generate \g-rays. The models are listed in Table~\ref{tcb:table-HDmodels}. Their key features are a considerably denser EDE (RUN03), higher explosion energy, and denser EDE (RUN10), or the absence of the accretion disk (RUN14).

The differences in the blast wave interaction with the environments may be seen in Fig.~\ref{tcorbor:fig-pp-images-models} where the \g-ray images of the remnant from each model are shown. The time moment for each snapshot is shown in the upper-right corner. Note that the remnant in each model is shown in about the same evolutionary phase as the remnant in the RUN04 model. In particular, the upper row corresponds to the time when the shock is within but near the edge of the accretion disk (if present in the model). The second row marks the situation shortly after the shock has completely overrun the RG star. 

Note that the evolutionary time-scale differs significantly in the RUN10 model. This is due to the higher explosion energy, which results in a substantially faster shock. Consequently, the remnant reaches the same size earlier. Another consequence is a higher $E\rs{max}$ and increased fluxes (the latter is also due to the higher EDE density). 
Differences in structure are also evident across runs. In particular, the structure around the RG resembling a bow shock is most prominent in RUN14 because there is no accretion disk, which introduces additional complexity into the system. In contrast, the quite dense EDE gradually becomes dominant in the hadronic \g-ray image in RUN03 (the last two images) and RUN10 (the last image). This feature is less prominent in RUN04 and RUN14 because in these models the EDE density is considerably lower and its extension in the $z=0$ plane is two times smaller. 

The next figures show how differences in interaction with the ambient medium are reflected in \g-ray spectra (Fig.~\ref{tcorbor:fig-pp-gamma-ray-spectra-models}) and light curves (Fig.~\ref{tcorbor:fig-pp-gamma-ray-lc-models}).

Initially, the shock speed in the RUN03 model is the same as in the reference model RUN04 because the outburst energy is the same. However, the densities of the CBM and disk are 100 times higher. Therefore, the shock decelerates faster. The lower velocity leads to a reduction of $\delta B$ and $E\rs{max}$. However, the higher density compensates for these losses to the level that \g-ray spectra in RUN03 (top plot in Fig.~\ref{tcorbor:fig-pp-gamma-ray-spectra-models}) are quite similar to those in RUN04. The light curves in the RUN03 model closely resemble those of RUN04 during a third of the first day. Then, the light curves in RUN03 decrease in a slightly different manner; namely, they deviate from a nearly power-law decay of emission in RUN04. 

As a result of the faster shock in model RUN10, the duration of the initial plateau in light curves is reduced by approximately a factor of two: the shock breaks out of the accretion disk faster than in the reference RUN04. The amplitudes of the \g-ray spectra are almost two orders of magnitude larger than the light curves in this model.

The light curves for RUN04 and RUN14 are very similar after the first eight hours (0.3 days). At earlier times ($t<0.3\un{day}$), the differences between them are due to the absence of an accretion disk in the model RUN14. The absence of the disk leads to lower $E\rs{max}$ and lower normalization of the initial \g-ray spectra.

In terms of contribution to the light curves, they are similar to the reference model. Our calculations show that the ejecta dominates the early phase in all models. The contribution from the disk to the \g-ray flux is lower, and that from the shocked CBM is lower than that. The shocked CBM, especially the EDE, becomes dominant in GeV photons after approximately 10 days in RUN03, 1 month in RUN10, and 3 months in RUN14.

\begin{table}
\caption{Number of {\it muonic} neutrinos which potentially could be detected from \tcb within the first four hours, 
as calculated with Eq.~(\ref{tcb:number-neutrino}) for our models with $\mu\neq 0$.}
\begin{tabular}{cccc}
\hline
Model & KM3NeT & IceCube    & IceCube \\
      & ARCA & Main Array & DeepCore \\
      & $10^3$-$10^8\un{GeV}$ & $10^{2.5}$-$10^9\un{GeV}$ & $10$-$10^4\un{GeV}$ \\
\hline
RUN04 & 2.2 & 0.8 & 1.9 \\
RUN03 & 5.3 & 2.0 & 3.9 \\
RUN10 & 64 & 24 & 40 \\
RUN14 & 0.0 & 0.0 & 0.0 \\
\hline
\end{tabular}
\label{tcb:table-neutrino}
\end{table}

\subsection{Detectability of neutrino emission}


In order to have an idea about the number of neutrino detections from the \tcb nova event, we approximate the effective area for muonic neutrinos in KM3NeT ARCA 
\citep[figure~7a][]{2024EPJC...84..885K} in the range $10^3-10^8\un{GeV}$:%
\begin{equation}
 A\rs{eff}^{\nu_\mu}=2\cdot \left[0.20\,(E/E\rs{0})^{-0.51}+0.46\,(E/E\rs{0})^{-0.06}\right]^{-6.4}\un{m^2}
 \label{tcb:km3app}
\end{equation} 
where the neutrino energy $E$ is in GeV, $E\rs{0}=10^4\un{GeV}$.
The approximation for the effective area of the IceCube Main Array \citep[blue line  on figure~1 in][]{2017ApJ...835..151A} for muonic neutrinos at energies $10^{2.5}-10^9\un{GeV}$ is 
\begin{equation}
 A\rs{eff}^{\nu_\mu}=\left[0.25\,(E/E\rs{0})^{-0.5}+0.45\,(E/E\rs{0})^{-0.04}\right]^{-6.5}\un{m^2}
 \label{tcb:ic1app}
\end{equation} 
with $E\rs{0}=10^4\un{GeV}$, 
and for the IceCube DeepCore \citep[figures~8-9 in][]{2012APh....35..615A} in the range $10-10^4\un{GeV}$ is 
\begin{equation}
 A\rs{eff}^{\nu_\mu}=\left[500\,(E/E\rs{0})^{-4.2}+33\,(E/E\rs{0})^{-1.4}\right]^{-1}\un{m^2}
 \label{tcb:ic2app}
\end{equation}
with $E\rs{0}=10^2\un{GeV}$.
Then, one can estimate the number of muonic neutrinos that may eventually be detected as
\begin{equation}
 N\rs{\nu_\mu}= \frac{c\rs{\nu}}{4\pi d^2}\int\int dN\rs{\nu}(E,t)/dE\cdot A\rs{eff}^{\nu_\mu}(E)\,dE\, dt
 \label{tcb:number-neutrino}
\end{equation}
where the integral over $dt$ corresponds to a duration of observations (taken here to be the first four hours), $dN\rs{\nu}/dE$ the total neutrino spectrum at the source ($\nu\rs{\mu}$, $\nu\rs{e}$ and their antiparticles) which we calculate from our 3D models, the coefficient $c\rs{\nu}=1/3$ accounts for neutrino oscillations (equal proportions of $\nu\rs{\mu}$, $\nu\rs{e}$ and $\nu\rs{\tau}$ at the Earth). 

To estimate the number of all-flavor neutrinos, it is necessary to multiply the integrand in (\ref{tcb:number-neutrino}) by a scaling function $F(E)=1+A\rs{eff}^{\nu_e}/A\rs{eff}^{\nu_\mu}+A\rs{eff}^{\nu_\tau}/A\rs{eff}^{\nu_\mu}$ where $A\rs{eff}^{\nu_i}$ are effective areas for neutrinos of different flavors. These effective areas depend on the neutrino energy, declination, and other factors, e.g. \citep{2024EPJC...84..885K,2021PhRvD.104b2002A,2017ApJ...835..151A}. Typically, 
$F$ takes values between 1 and 3 at a given declination, but can be larger at some neutrino energies, e.g., due to the Glashow resonance.

The number of muonic neutrinos estimated with Eq.~(\ref{tcb:number-neutrino}) is shown in Table~\ref{tcb:table-neutrino} for our four HD models. It appears that the best chance of detecting neutrinos from \tcb with KM3NeT and IceCube is for the RUN10 model, i.e., if the energy of the outburst is above $E\rs{bw}=10^{44}\un{erg}$, which is rather an extreme assumption. Models RUN04 and RUN03 predict an order of magnitude lower neutrino flux compared to RUN10, even though the density of target protons is rather high in these models. If there is no accretion disk (model RUN14) or the disk has a low density, then \tcb cannot produce enough neutrinos to be detected either by KM3NeT or IceCube.

Furthermore, \tcb is not in a favorable position for KM3NeT, as both are located in the northern hemisphere. The detectability of \tcb by KM3NeT/ARCA should even depend on the time of day. With a declination of $+25\degr 55'$, \tcb is below the horizon at the location of KM3NeT/ARCA for only part of the day.
IceCube, being in Antarctica, is in a more favorable position to detect neutrinos from \tcb outburst. 

\subsection{Possible effects from the photon field of the nova} 
\label{tcorbor:sec-nova-radiation}

\subsubsection{Enhanced Inverse-Compton gamma-rays}

As evident from the light curve of \tcb from 1866, its apparent magnitude changed from the maximum 2.0 to 3.0 during the first day \citep{1946PASP...58..153P}. This corresponds to bolometric luminosities $L\sim 10^{38}\un{erg/s}$ (Appendix~\ref{tcb:app-opticlc}). 
Taking the shock radii $R\sim 0.5\un{a.u.}$, we may estimate the average radiation energy density $\omega\rs{ph}=3L/(4\pi R^2c)\sim 10\un{erg/cm^3}\sim 10^{13}\un{eV/cm^3}$. Such a powerful photon field could result in a leptonic \g-ray emission which is compatible with the hadronic one (this could be seen, e.g., from Fig.~\ref{tcorbor:fig-pp-specta} by multiplying the long-dashed lines with a factor $\sim 10^{13}$).

\subsubsection{Absorption of gamma-rays}
\label{tcorbor:sect-absorp}

In an opaque medium, the intensity is $I=I_0\exp(-\tau)$ where $I_0$ is the initial intensity, the optical depth $\tau$ is an integral of the absorption coefficient $\mu$ over a distance ${\cal L}$; in a uniform medium $\tau=\mu {\cal L}$.  
The flux of GeV-PeV \g-rays could be reduced by the interaction of \g-photons with matter or less energetic photons. Such interactions typically produce $e^{+}+e^{-}$ pairs. 

For \g-rays with energy $0.5-100\un{GeV}$, the absorption coefficient is caused by interactions with ionized plasma reaching hydrogen atoms \citep[][p.238]{1990acr..book.....B}: 
\begin{equation}
 \mu\rs{\gamma p}(E\rs{\gamma})\simeq 3.6\E{-27}n\rs{H}\left[\ln(E\rs{\gamma}/m\rs{e}c^2)-1.9\right]\ \un{cm^{-1}}
\end{equation}
where $n\rs{H}$ is the hydrogen number density in $\un{cm^{-3}}$. 
The density within the accretion disk is $n\rs{H}\simeq 10^3n\rs{ede}=(10^9-10^{11})\un{cm^{-3}}$ depending on a model. 
The optical depth in the disk for $\sim 10^{9}\un{eV}$ photons is 
$\tau\simeq 0.025$ for RUN04, $0.25$ for RUN10 and $2.5$ for RUN03. Outside the accretion disk, $\tau$ is three orders of magnitude lower for each model of \tcb. 
Therefore, the \g-ray spectrum and the light curve in the RUN03 model may be affected by absorption of GeV \g-rays within the accretion disk. 
This should be accounted for in the modeling of \tcb if observations of the nova provide evidence about accretion disk density $\gtrsim 10^{11}\un{cm^{-3}}$. 
For reference, the shock runs out of the disk in the RUN03 model in about 6 hours.

Another mechanism for the absorption is the interaction with the photon field. Then $\mu\rs{\gamma\gamma}\simeq \sigma\rs{\gamma\gamma}\omega\rs{ph}/\epsilon$ where $\sigma\rs{\gamma\gamma}$ is the cross section, $\omega\rs{ph}$ is the energy density of the background photons with energy $\epsilon$. 
The maximum of the cross-section occurs for \g-rays with energy $E\rs{\gamma}=0.75\epsilon\rs{eV}^{-1}\un{TeV}$ where $\epsilon\rs{eV}$ is $\epsilon$ in eV \citep{1983Ap.....19..187A}. The maximum of black-body radiation with $T\rs{eff}\simeq 10^4\un{K}$ is at $\epsilon\approx2.4\un{eV}$. Thus, this kind of absorption affects TeV \g-rays. 

For interactions of \g-rays with such optical photons, the maximum coefficient is $\mu\rs{\gamma\gamma}\simeq 7\E{-26}\omega\rs{ph}\un{cm^{-1}}$ 
for \g-rays with energy $\sim10^{12}\un{eV}$ 
where $\omega\rs{ph}$ is the radiation energy density in $\un{eV/cm^3}$ \citep{1990acr..book.....B}. 
A remnant's radius about $1\un{a.u.}$ and an optical photon field with $\omega\rs{ph}\sim 1\un{eV/cm^3}$ result in $\tau\sim 10^{-12}$
, that is negligible. 
Instead, the energy density of optical photons from the nova itself of the order $\omega\rs{ph}\sim 10\un{erg/cm^3}$ could be responsible for the optical depth $\tau\simeq \mu\rs{\gamma\gamma}R \sim3$. Therefore, the absorption due to the interaction of \g-rays with optical photons of the nuclear-burning WD may be important during the first days after the outburst.

\subsubsection{Eventual effects}
\label{tcorbor:sect-absorpeffects}

A radiation field from nuclear-burning WD could affect the \g-ray emission through an enhancement of leptonic radiation and \g-ray absorption. Instead, it does not affect the neutrino flux. 

In order to explore a framework for the analysis of future observations, we synthesize the observables including relevant IC emission and \g-ray opacity.  
For this purpose, we reconstruct the evolution of the \tcb luminosity from observations of the 1946 outburst and adopt a spatial structure for the optical photons as described in Appendix~\ref{tcb:app-opticlc}.

\begin{figure}
  \centering 
  \includegraphics[width=0.99\columnwidth]{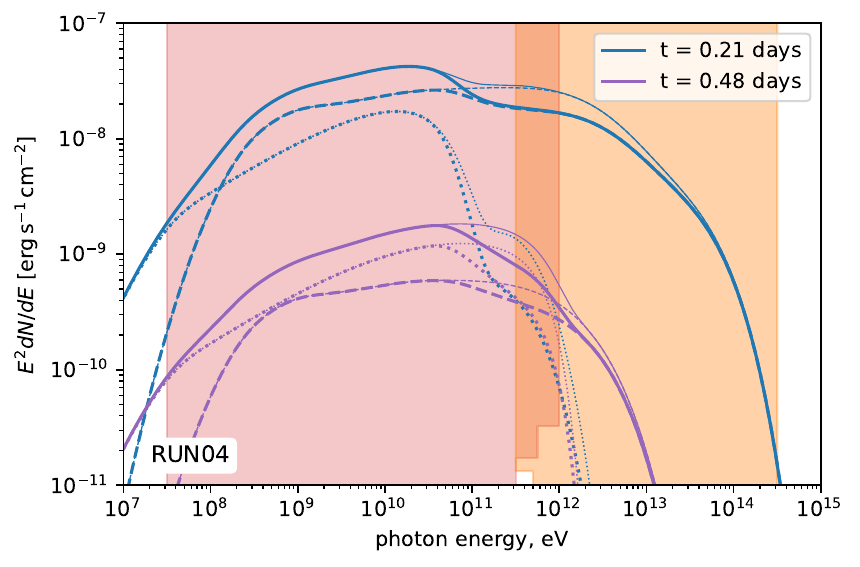} 
  
  \caption{%
    Gamma-ray spectra for the RUN04 model with $\mu \ne 0$ for two time moments. 
    Solid lines correspond to the total spectrum, given by the sum of the hadronic (dashed) and leptonic (dotted) components. Thick lines include absorption, while thin lines do not. The photon field temperature is $10^4\un{K}$. The shaded areas correspond to the Fermi and ASTRI energy ranges. 
    The thin dashed blue line is the same as the solid blue line in Fig.~\ref{tcorbor:fig-pp-specta}.
  }
  \label{tcorbor:fig-ic-pp-absorption}
\end{figure}

\begin{figure}
  \centering 
  \includegraphics[width=0.97\columnwidth]{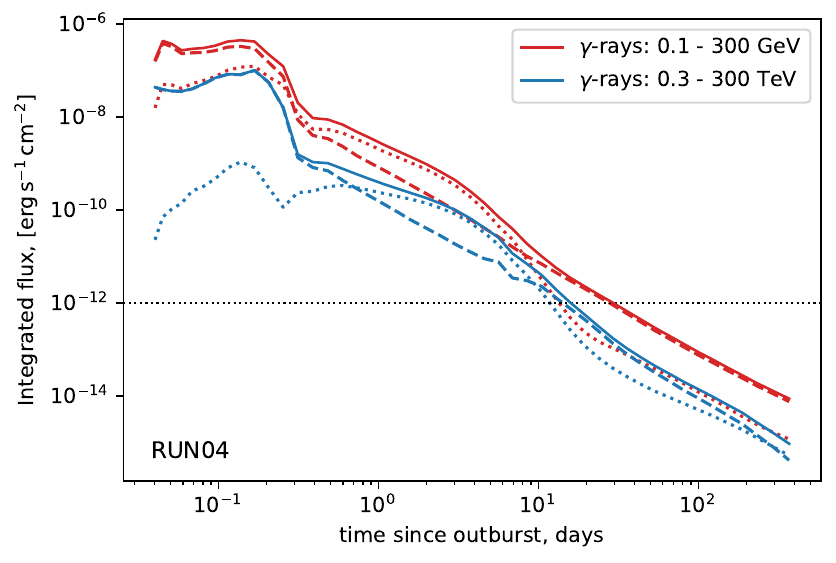} 
  \caption{%
       Gamma-ray light curves for the model RUN04 with $\mu \ne 0$, the photon field temperature $10^4\un{K}$, and absorption included. Red lines correspond to the Fermi range, and blue lines correspond to the ASTRI range. Dashed and dotted lines represent the hadronic and leptonic (IC) components; their sum is shown by the solid lines.
  }
  \label{tcorbor:fig-ic-pp-absorption-LC}
\end{figure}

The \g-ray spectra are shown in Fig.~\ref{tcorbor:fig-ic-pp-absorption}. The effect of absorption is visible from the comparison of the thick and thin lines: it mostly affects TeV \g-rays, as expected. The dominant contribution to the total \g-radiation varies; it could be either from hadronic (as at $t=0.21\un{days}$) or IC ($t=0.48\un{days}$) processes. This is also visible from the light curves in Fig.~\ref{tcorbor:fig-ic-pp-absorption-LC}. The hadronic emission is expected to be dominant during the first day and then after about one week. Instead, the leptonic emission could be dominant during the first week. 

It is expected, and our calculations confirm, that a higher (lower) photon-field temperature results in increased (decreased) IC emission and \g-ray absorption which affects \g-rays with higher (lower) energies.

\section{Conclusions}
\label{tcorbor:sec-concl}

We explore the potential \g-ray and neutrino production in the remnant of an expected 
\tcb nova. To this end, we synthesize the \g-ray and neutrino emission from several 3D HD models of \tcb. 
In this way, we explore assumptions about the structure of the medium surrounding the 
\tcb system and the energy of the nova event. This will allow us to constrain the parameter space and model the early stages of shock-wave propagation following the actual nova outburst, guided by future observations.

In the present paper, we explore for the first time the effect of non-uniformity of the downstream flow on CR acceleration in novae and compare it to a common assumption of the flow speed uniformity. 
We find that accounting for flow non-uniformity yields higher maximum energies of CRs, which create more favorable conditions for the observation of TeV photons and neutrinos.

Our main conclusions are the following.
\begin{itemize}
\item \tcb should be detectable for about one month in GeV photons and about a week in the TeV band. 
\item \tcb should be detectable by neutrino experiments, unless the density of the accretion disk in the \tcb system is low, $\lesssim n\rs{ede}\sim 10^{7}\un{cm^3}$.
\item The probability of detecting neutrinos from \tcb is highest in the RUN10 model with the highest explosion energy, $10^{44}\un{erg}$.
\item During the period of detectability, the dominant contribution to the hadronic \g-ray and neutrino fluxes is from ejecta illuminated by the protons accelerated at the shock. Contributions to the \g-ray flux from the illuminated shocked and ambient CBM are negligible compared to the ejecta during the first months. 
\item Contribution to the hadronic \g-ray flux from the shocked CBM material could dominate the \g-rays after a few months, but by then the emission from \tcb could not be detected anymore by modern experiments. 
\item During the first days after the outburst, the strong radiation field from the nuclear-burning white dwarf could result in efficient IC \g-radiation and prominent absorption of \g-rays.
\item Strong negative gradients in the flow speed can lead to increased acceleration efficiency, i.e., to more efficient transfer of energy into CRs. 
\item With this effect, the actual $E\rs{max}$ could be an order of magnitude larger than given by formulae in Sect.~\ref{tcb:sect-pmax}.
\item Ejecta protrusions provide conditions for the most efficient cosmic-ray acceleration due to the strong non-uniformity of the plasma velocity; in these regions, the downstream flow speed increases with distance from the shock and becomes larger than the immediate post-shock speed.
\item Similar conditions occur when the forward shock encounters the dense material of the RG star: the shock quickly decelerates, while the plasma behind it continues to flow because it does not immediately respond to the sharp density increase.
\item We reach the surprising finding that even with lower energy than supernovae, novae can accelerate particles to almost PeV energies, though in models with very high explosion energy, high densities, and high amplified magnetic fields (like in the RUN10 model).
\item Our simulations demonstrate that the 3D structure of the nova remnant results in significant spatial variation in maximum particle energy and the spectrum normalization across the shock surface, particularly in the vicinity of the EDE and accretion disk. This could lead to different features in the integrated \g-ray spectra, like spectral breaks and bumps. 
\item While our multi-zone approach captures the complex geometry, the absolute normalization of the predicted fluxes remains sensitive to the assumed injection and acceleration efficiencies, especially to their temporal evolution. Future observations of the \tcb event will be essential to calibrate these microphysical parameters.
\end{itemize}

The correlation between the early-phase \g-ray and the thermal X-ray emission (studied in Paper I) provides a robust diagnostic for the nova outburst properties, the shock velocity evolution, and the structure of the circumbinary environment.

Our models suggest that multi-messenger observations should prioritize the first 24 hours to capture the disk-interaction phase, which is critical for distinguishing between different CBM configurations. 

{\citet{2025arXiv250707096T}, by simply scaling the expected neutrino flux from the optical brightness of the nova and comparing it with observations of RS~Ophiuchi, estimated that \tcb is unlikely to be detected by IceCube (their figure 6). Our estimates, based on 3D HD simulations incorporating post-shock velocity gradients and the detailed structure of the CBM, predict higher maximum neutrino energies and a stronger signal due to shock propagation through the dense accretion disk, making neutrino detection more probable.

Finally, while this study focuses on the specific environmental parameters of \tcb, our modeling framework—combining 3D hydrodynamics with diffusive shock acceleration, which accounts for the non-uniformity of the flow structure on the CR acceleration scales—offers a powerful tool for evaluating the high-energy messengers from similar systems. More specifically, extending such an analysis to observations of other recurrent novae and more energetic systems, such as interacting supernovae (e.g., SN 2014c \citet{2024ApJ...977..118O}), could be essential for determining their overall contribution to the population of Galactic cosmic rays. In particular, high initial shock velocities, quite dense environment in the vicinity of a progenitor due to previous mass loss activity, and, as a result, essential enhancement of the turbulent MF during the early phases of (super)nova could naturally result in CRs with energies around a knee in the CR spectrum.

\begin{acknowledgements}
We are grateful to Ileana Chinnici for drawing our attention to the fact that \tcb was the first nova observed spectroscopically. 
We acknowledge the support from INAF 2023 RSN4 Theory grant.  L.C. is grateful for support from NSF grant AST-2107070 and NASA grants 80NSSC23K0497 and 80NSSC25K7334.
F.B. and S.U. acknowledge support from KM3NeT4RR of Project National Recovery and Resilience Plan (NRRP), Mission 4 Component 2, Investment 3.1, Funded by the European Union – NextGenerationEU, Concession Decree MUR No. n. Prot. 123 del 21/06/2022. F.B. acknowledges partial support from INAF Grant AF2024.
O.P. and T.K. thank the Armed Forces of Ukraine for providing security to perform this work. 
This research used the HPC system MEUSA at the SCAN (Sistema di Calcolo per l’Astrofisica Numerica) facility for HPC at INAF-Osservatorio Astronomico di Palermo, Italy.
\end{acknowledgements}

\bibliographystyle{aa}
\bibliography{tcorbor} 

\begin{appendix}  

\section{How the gradient of the flow speed is calculated}
\label{tcorbor:app-mu}

The flow speed gradient is calculated using the directional derivative along the flow lines: $\nabla_\mathbf{v} v = \mathbf{\hat{v}} \cdot \nabla v$, where $\mathbf{\hat{v}}$ is the unit vector along a velocity field line (Fig.~\ref{tcorbor:fig-3d-volume-slice}). Numerically, we computed it using a second-order backward scheme with the origin at the shock front surface. 
Our simulations are performed on a uniform $512^3$ grid. Therefore, the 3 pixels involved in the calculation of the derivative correspond to about 1-2 percent of the shock radius. 

Although the ranges of $\mu$ and $E\rs{max}$ could be quite wide, most cells have values concentrated within narrower intervals. for example, Fig.~\ref{tcorbor:fig-2d-hist-mu-Emax} shows a distribution of values $(\mu, E\rs{max})$ on a shock surface in the RUN04 model on day $153$. The color scale shows the number of grid zones on the remnant's surface with given values of these parameters. The biggest contribution to the overall particle spectrum is from points that are clustered between $\mu\approx -20\div 20$ and $E\rs{max}\approx 100\div700\un{GeV}$. 

For each dataset (i.e., for a given model at a given time), we apply the data cleaning procedure. We calculate the mean $\langle\mu\rangle$ and the standard deviation $\sigma\rs{\mu}$ across the shock and exclude from consideration the outliers, i.e., cells where the value of $\mu$ is beyond the range $\langle\mu\rangle\pm3\sigma\rs{\mu}$. The fraction of such points is typically $\lesssim1\%$. 
We also noticed that at early times, there could be cells with extremely high absolute values of $\mu$, which considerably affect the standard deviation. We treat such values as numerical artifacts. Therefore, in the second step, we exclude the cells with $\mu$ out of the $(-100,100)$ range. This could increase the fraction of rejected cells to a few percent. We stress that, at any time, most cells have absolute values of $\mu$ less than $100$ (Fig.~\ref{tcorbor:fig-hist-mu-values}).

\begin{figure}[!h]
  \centering 
  \includegraphics[width=\columnwidth]{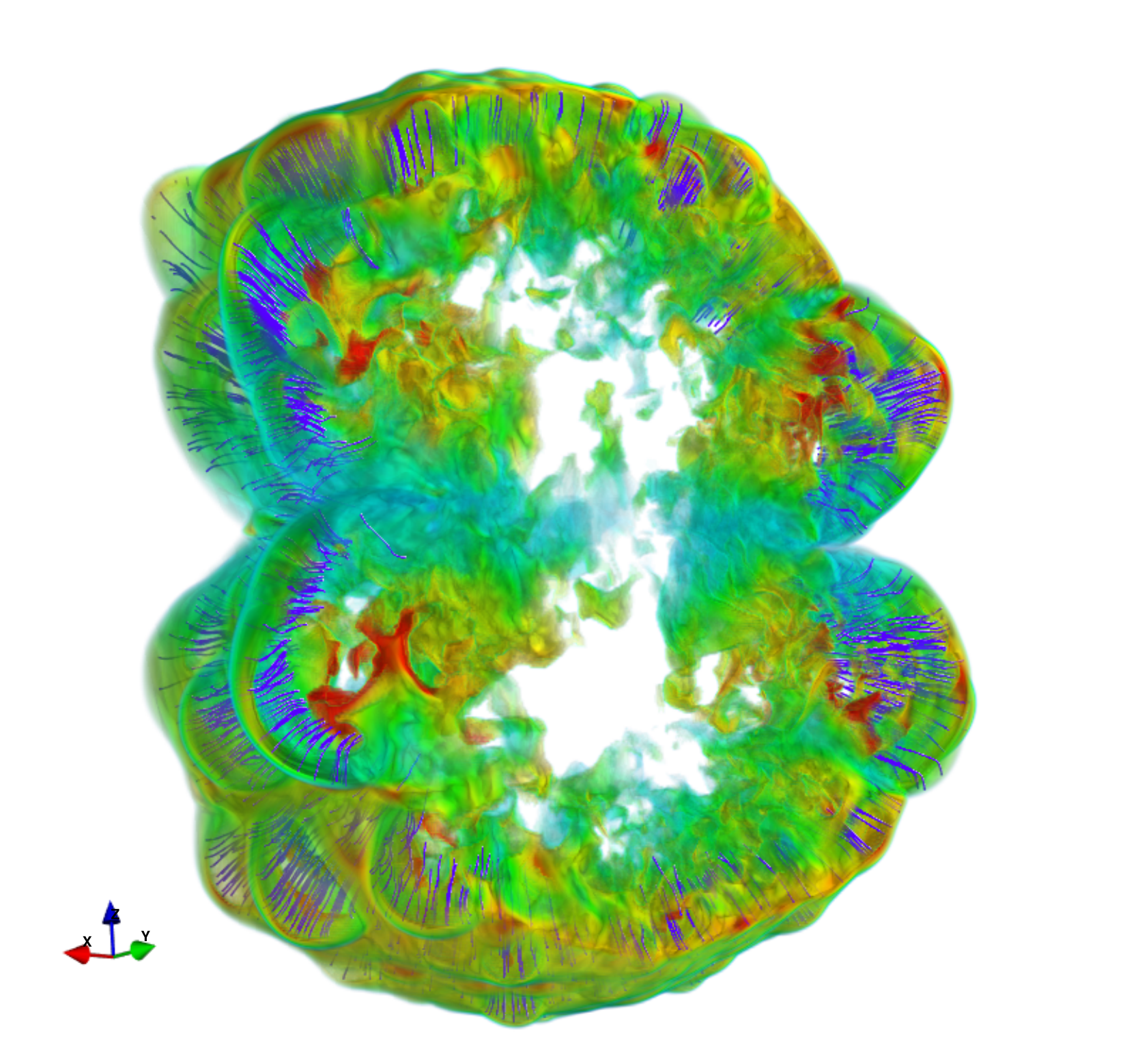}
  \caption{%
       A slice of the 3D volume rendering of the flow velocity. Blue lines indicate flow streamlines backtraced from the shock front. Model RUN04, $t=153\un{days}$.%
  }
  \label{tcorbor:fig-3d-volume-slice}
\end{figure}
\begin{figure}
  \centering 
  \includegraphics[width=\columnwidth]{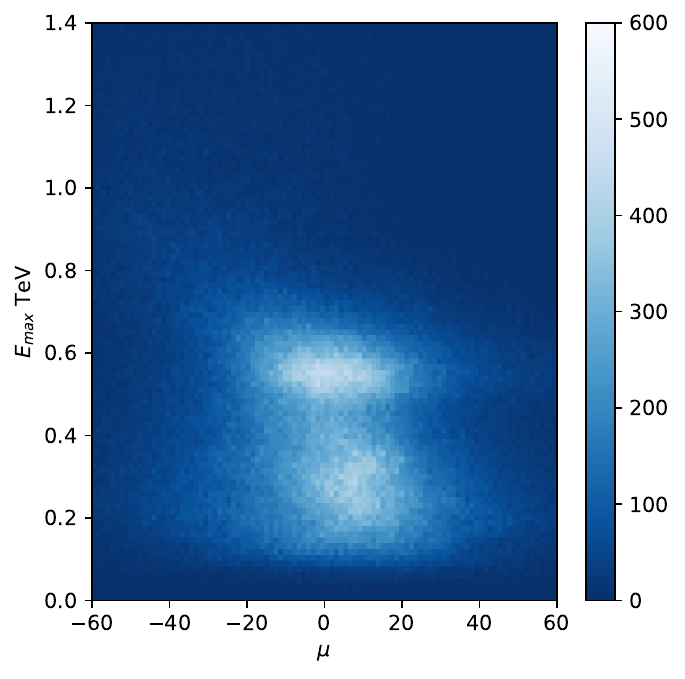}
  \caption{%
       The distribution of the values $(\mu, E\rs{max})$ for pixels across the shock surface. $E\rs{max}=\mathrm{min}(E\rs{m1},E\rs{m2})$ for protons. The color scale shows the number of cells. Model RUN04, $t=153\un{days}$.%
  }
  \label{tcorbor:fig-2d-hist-mu-Emax}
\end{figure}
\begin{figure}
  \centering 
  \includegraphics[width=\columnwidth]{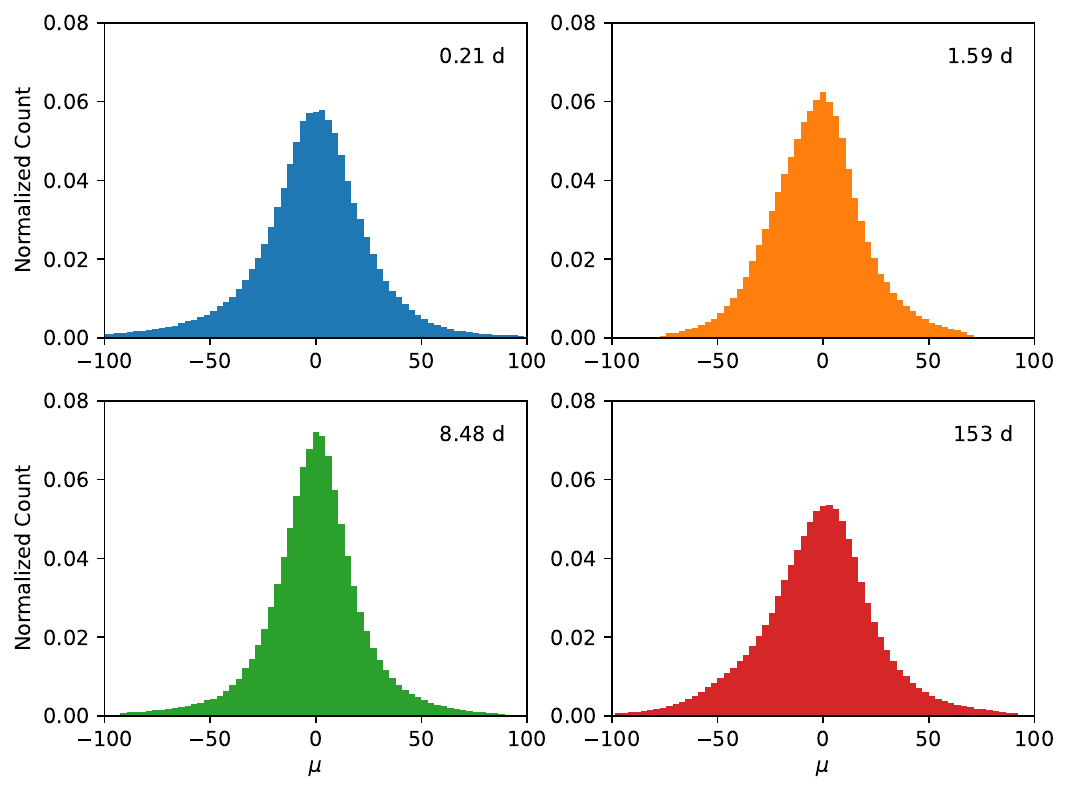} 
  \caption{%
      Histograms of the values of $\mu$ at the shock front for RUN04 at different times. The counts are normalized by the total number of grid cells on the shock front after the data cleaning. Values are shown after the cells with values outside the $\pm 3\sigma$ and (-100,100) intervals are excluded. 
  }
  \label{tcorbor:fig-hist-mu-values}
\end{figure}

Note that the negative gradient of the flow speed downstream of the shock leads to an increased acceleration efficiency. Physically, this occurs because the material behind the shock moves faster than that immediately post-shock. Then the shock compression factor, which is 'visible' to CRs, could be larger than the standard $4$. This effect can arise, for example, in regions with ejecta protrusions or when the shock encounters a strong density enhancement. As a result, the CR energy could reach a substantial fraction of the shock's kinetic energy, leading to a non-adiabatic behavior of nova and supernova shocks at very early times after the explosion. The loss of adiabaticity is then driven by efficient energy transfer to CRs rather than by radiative losses. These considerations emphasize the need for a dedicated study of this effect.

\begin{figure*}
  \centering 
  \includegraphics[width=\textwidth]{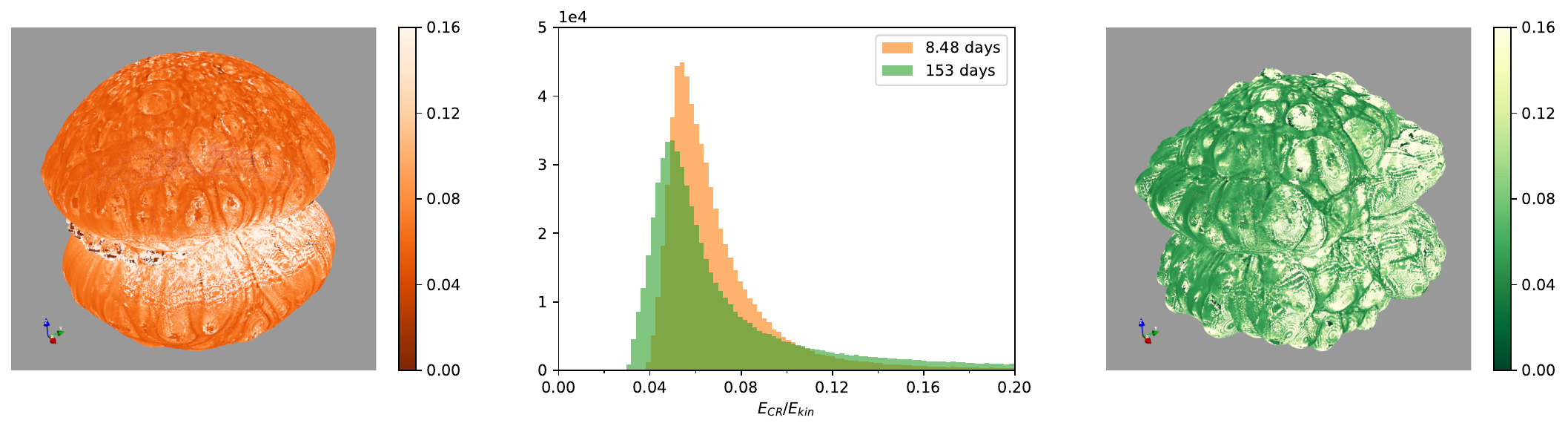}
  \caption{%
       Spatial distributions and histograms for the fraction $\xi\rs{cr}$ of kinetic energy of the shock transferred to CRs. It is calculated by integration of local CR spectra and local shock kinetic energy. HD model considered is RUN04 at $t=8.5\un{day}$ and $153\un{day}$. 
       The average acceleration efficiency across the shock surface, calculated from the average CR spectra at these times (the spectra are shown by solid lines in Fig.~\ref{tcorbor:fig-proton-specta}), is $\hat\xi\rs{cr}=0.08$ and $\hat\xi\rs{cr}=0.1$ respectively. These plots are calculated by keeping the injection efficiency $\xi\rs{in}$ constant in time and space. 
  }
  \label{tcorbor:fig-hist-Ekin-Ecr-Efrac}
\end{figure*}

\section{Note about the acceleration and injection efficiencies}
\label{tcorbor:app-xicr}

The momentum distributions of protons and electrons depend on the injection efficiency $\xi\rs{in}$ and the acceleration efficiency $\xi\rs{cr}$. 
These parameters are typically set to some numbers and kept constant.
However, they are not independent, even in a one-zone model. In particular, the energy of CRs, given by the integral $E\rs{cr}=\int E N(E)dE$, depends on the amplitude $K$ of the CR distribution $N(E)$, which is affected by $\xi\rs{in}$ (see Eq.~\ref{tcb:defK}). 
Therefore, one cannot assume that both efficiencies are constant. 

The situation is more complex in a nonstationary multi-zone model where parameters evolve in time and vary across the shock. Fig.~\ref{tcorbor:fig-hist-Ekin-Ecr-Efrac} illustrates this. It shows the distribution of local acceleration efficiencies $\xi\rs{cr}$ across the shock surface for two time moments. To obtain the local fraction of the shock kinetic energy converted into CRs $\xi\rs{cr}$, we integrate the local CR spectrum and divide the resulting CR energy by the kinetic energy of the shock in that location. This is {\it a posteriori} calculation of the acceleration efficiency. However, to calculate the CR spectrum, we need to set $\xi\rs{cr}$ {\it a priory} because it affects the level of amplified MF $\delta B\propto \xi\rs{cr}^{1/2}$ and thus the diffusion properties $D\propto \delta B^{-1}$ and possible values of maximum energy $p\rs{m1}\propto \delta B$ as well as $p\rs{m3}\propto \delta B$ of protons and $p\rs{m3}\propto \delta B^{-1/2}$ of electrons; the second option for the maximum momentum depends on the acceleration efficiency directly $p\rs{m2}\propto \xi\rs{cr}$  (Sect.~\ref{tcorbor:sec-method}). 

We also consider a parameter characterizing the overall shock, namely, the average acceleration efficiency $\hat\xi\rs{cr}$. To evaluate it, we integrate the average CR spectrum (i.e., averaged over the shock surface) and divide the resulting CR energy by the average shock kinetic energy. 

One may assume only one parameter constant, either injection or acceleration efficiency.
To produce Fig.~\ref{tcorbor:fig-hist-Ekin-Ecr-Efrac}, we kept {\it the injection efficiency $\xi\rs{in}$ constant} in time and space. Its value is set to provide the {\it a posteriori} average acceleration efficiency $\hat\xi\rs{cr}=0.10$ at $t=153\un{day}$. We also set the {\it a priori} (because we need it to initiate calculations of the CR spectra) local values $\xi\rs{cr}=0.1$ to be constant in space and time while evaluating $\delta B$ and $p\rs{m2}$ from Eqs.~(\ref{tcorbor:eq-dB}) and (\ref{tcorbor:eq-pm2}). As a result, the average acceleration efficiency $\hat\xi\rs{cr}$ varies with time. Namely, $\hat\xi\rs{cr}=0.04$, $0.07$, $0.1$ at $t=1.6,\ 8.5,\ 153\un{days}$, respectively. It increases by a factor of $2.5$ between days $1.6$ and $153$. To this end, there is a contradiction between the {\it a priori} and {\it a posteriori} values of acceleration efficiency. The reason for this discrepancy is the constancy of the injection efficiency $\xi\rs{in}$. 

There are two ways to resolve this contradiction. First, one may assume the injection efficiency $\xi\rs{in}$ is the same for any $t$. Then one needs to adopt an iterative technique: assume some {\it a priori} $\xi\rs{cr}$, calculate the CR spectrum, its energy, and then {\it a posteriori} $\xi\rs{cr}$. If the two values differ, the initial value has to be updated, and the procedure to be repeated until convergence is achieved within the required accuracy. Such a procedure requires many iterative steps because $\xi\rs{cr}$ affects both the CR normalization $K$ and the maximum energy $p\rs{max}$. The acceleration efficiency varies over time with this procedure. 

In the second approach, one may assume {\it a constant acceleration efficiency $\xi\rs{cr}$}. Then, for each $t$, the {\it a priori} and {\it a posteriori} $\xi\rs{cr}$ may be equilibrated by simply changing the injection efficiency $\xi\rs{in}$ because it affects the normalization $K$ only. Under such an approach, the injection efficiency varies over time. However, it requires only one additional iterative step after calculation of the CR spectrum, namely, a renormalization of $\xi\rs{in}$. In our paper, we follow this second strategy because it is numerically more appropriate for 3D multi-zone modeling. 

A similar ambiguity arises from the spatial variation of $\xi\rs{cr}$ at a fixed moment of time. 
An example of distributions of local values of $E\rs{cr}/E\rs{kin}$ across the shock surface is shown in Fig.~\ref{tcorbor:fig-hist-Ekin-Ecr-Efrac}. 
In principle, one should apply the methodology described above to each cell across the shock. However, we use the average value $\hat\xi\rs{cr}$ in Eqs.~(\ref{tcorbor:eq-dB}) and (\ref{tcorbor:eq-pm2}) even locally ($\xi\rs{in}$ is then spatially constant, too). 
In this way, we simulate a spatial variation of the local {\it a posteriori} acceleration efficiency, which is more natural than assuming a strictly uniform efficiency everywhere.  
Moreover, this approach is supported by the fact that the diffusion of escaped particles is very fast, so we may assume that particles from the entire shock surface contribute to the local MF amplification. In other words, the local values of $\delta B$ depend on the CR spectrum produced everywhere over the entire shock surface. 

\begin{figure}[!h]
  \centering 
  \includegraphics[width=\columnwidth]{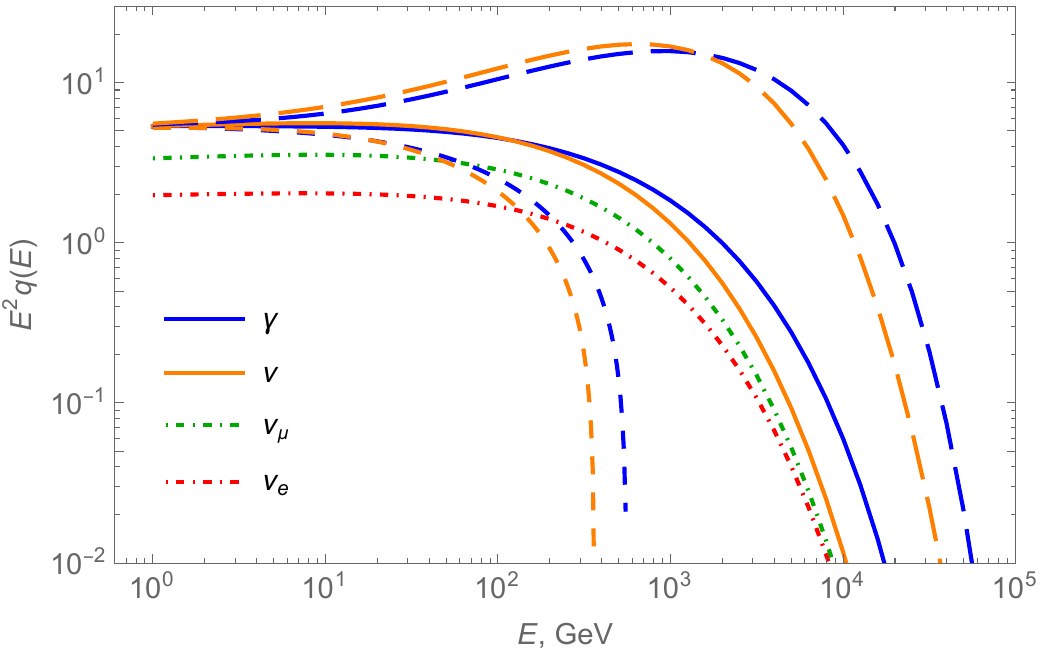}
  \caption{%
        Spectra of \g-rays (blue lines) and neutrinos (orange lines, which are sums of muonic and electronic neutrinos and antineutrinos) produced in $pp$ interactions of relativistic protons having the momentum distribution (\ref{tcb:eqN}) with protons at rest, in arbitrary units. Three pairs of lines represent the cases with $\mu=30$ (short dashed), $\mu=0$ (solid), $\mu=-30$ (long dashed). Other inputs are $p\rs{max}=10\un{TeV}$, $\bar x\rs{p}(p\rs{max})=0.1$, Bohm diffusion. 
        Contributions from the muonic (green dot-dashed line) and electronic (red dot-dashed line) neutrinos are shown for the case $\mu=0$.
  }
  \label{tcorbor:fig-app-ngspectra}
\end{figure}

\section{Comparison of the gamma-ray and neutrino spectra}
\label{tcorbor:app-1}

In order to compare the \g-ray and neutrino spectra from the proton-proton interactions, we calculate them as \citep{2006PhRvD..74c4018K}
\begin{equation}
	q(E)=cn\rs{H}\int_{0}^{1} \sigma\rs{pp}(E/x)F(x,E)N(E/x)\frac{dx}{x}
	\label{pp:ppintini}	
\end{equation}
where $x=E/E\rs{p}$, $E\rs{p}$ is the energy of incident protons, $E$ is the energy of products (\g-rays or neutrinos), $N(E\rs{p})$ is the energy distribution of protons, $\sigma\rs{pp}$ is the cross-section for the $pp$ collisions which we take from the work by \citet{2014PhRvD..90l3014K}, $F(x,E)$ is a spectrum of a product from a `single' $pp$ interaction. We use the parameterizations of $F(x,E)$ for \g-rays, electronic and muonic neutrinos developed by \citet{2006PhRvD..74c4018K}, which are valid for $E\geq 1\un{GeV}$. 

Fig.~\ref{tcorbor:fig-app-ngspectra} compares the spectra of \g-rays and neutrinos. The spectrum of emitted neutrinos (orange lines) follows the \g-ray spectrum (blue lines) quite closely except for the position of the high-energy cutoff. 
Therefore, a hadronic \g-ray spectrum produced with the Naima package could also serve as a reasonable proxy for the all-flavor neutrino spectrum, with a correction for a bit smaller maximum energy of neutrinos $E\rs{\nu,max}\approx0.8E\rs{\gamma,max}$.


\section{Photon field from the nuclear-burning WD}
\label{tcb:app-opticlc}

In order to calculate the radiation field of the \tcb nova, we adopt the light curve from the 1946 event \citep{1946PASP...58..153P,1946PASP...58Q.213P,2023MNRAS.524.3146S}. 
A typical procedure is followed to convert the visual apparent magnitude $m\rs{V}$ to the luminosity $L$. We adopted the distance $d=890\un{pc}$, bolometric correction $BC\rs{V}=-0.2\un{mag}$ and extinction in the visual band $A\rs{V}=0.2\un{mag}$. Novae typically peak at optic making $BC\rs{V}$ correction small; the value $BC\rs{V}$ we have chosen corresponds to the effective temperature $\approx 10^4\un{K}$ \citep{1996ApJ...469..355F,2010AJ....140.1158T}. The adopted value of extinction $A\rs{V}$ comes from the formula $R\rs{V}=A\rs{V}/E\rs{B-V}$ with a value of the extinction ratio $R\rs{V}\approx 3.1$ typical for ISM in our Galaxy \citep{1989ApJ...345..245C} and the color excess (reddening) between B and V bands $E\rs{B-V}\approx 0.07$ for \tcb \citep{2022NewA...9701859N}. 
Indeed, the reddening is small due to the proximity and high galactic latitude of \tcb.

Assuming the emission can be approximated by a black body, we keep the bolometric correction $BC\rs{V}$ and the effective temperature $T$ constant in time. There are evidences supporting this. The light curves in H and V bands follow one another closely with approximately constant $m\rs{H}-m\rs{V}$ in the recurrent nova U~Sco during 40 days after the outburst \citep[figure~1 in][]{2023MNRAS.522.4841E}. During the same period, the black body emission with the temperature $\sim10^4\un{K}$ is present in the spectrum of this nova (the same reference). 

The evolution of the luminosity derived in this way, is shown in Fig.~\ref{tcorbor:fig-app-lumin} together with its approximation
\begin{equation}
 L(t\rs{d})\approx 1.6\E{38}t\rs{d}^{-0.05}\exp\left(-t\rs{d}/3\right)+1.1\E{36}t\rs{d}^{-0.2}
 \label{tcorbor:app-eqLappr}
\end{equation}
where $t\rs{d}$ is in days since the explosion.

\begin{figure}
  \centering 
  \includegraphics[width=0.98\columnwidth]{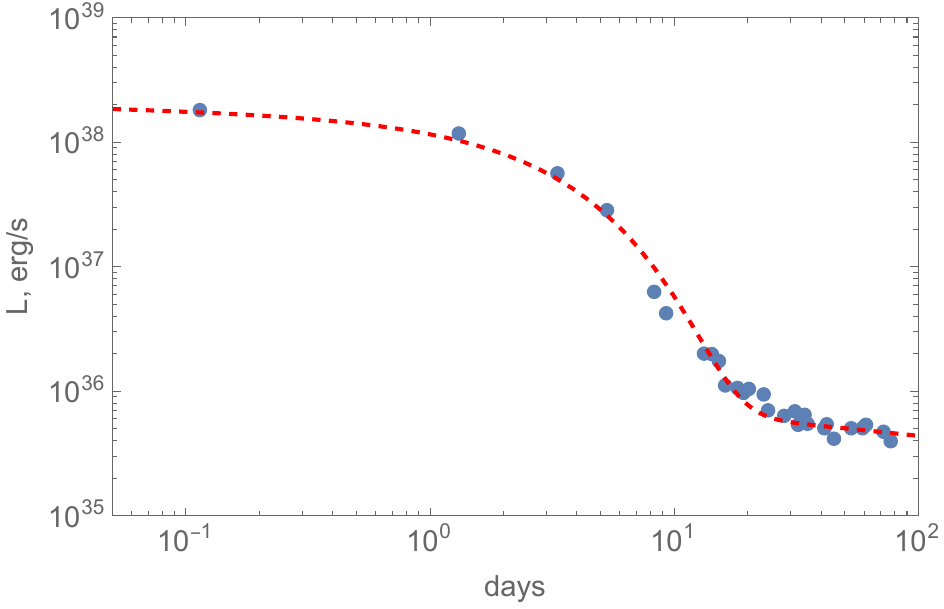}
  \caption{%
        Evolution of the \tcb luminosity during the 1946 outburst. Dots correspond to the observational data, the line shows the approximation (\ref{tcorbor:app-eqLappr}). 
  }
  \label{tcorbor:fig-app-lumin}
\end{figure}

The \g-ray emission in Sect.~\ref{tcorbor:sect-absorpeffects} is calculated as follows. 
The radiation energy density at a distance $r$ from WD is calculated as $\omega\rs{ph}(r,t)=L(t)/(4\pi r^2 c)$ where $c$ is the light speed. 
We set the effective temperature, which characterizes the black-body radiation, to $10^4\un{K}$. Assuming that the relative spatial variation of the effective temperature satisfies condition $\Delta T/T\ll 1$, the local black-body spectrum $\omega\rs{\epsilon,r}(\epsilon)$ varies in space at a given time $t$ in normalization:
\begin{equation}
 \omega\rs{\epsilon}(\epsilon,r,t)=\frac{\omega\rs{ph}(r,t)}{\omega\rs{0}}\omega\rs{0\epsilon}(\epsilon)
\end{equation}
where $\epsilon$ is the energy of the background black-body photons, $\omega\rs{0}$ is an integral of $\omega\rs{0\epsilon}$ over $\epsilon$, that is $\omega\rs{0}=aT^4$ with $a=7.58\E{-15}\un{erg/(cm^3\,K^4)}$. 

With local values of $\omega\rs{ph}$, we may calculate IC emissivity $\varepsilon_0$ in each grid cell. When the emitted (hadronic and leptonic) \g-radiation from a given cell $\varepsilon_0$ travels along the line of sight toward the observer, it is reduced due to interactions with the black-body photons to $\varepsilon=\varepsilon_0\exp(-\tau_{\gamma\gamma})$ where 
\begin{equation}
 \tau_{\gamma\gamma}(E\rs{\gamma},t)=\int\limits_{y_0}^{y_1} \int\limits_{\epsilon_0}^{\infty} 
 \sigma_{\gamma\gamma}(E\rs{\gamma},\epsilon)\ \omega\rs{\epsilon}(\epsilon;x_0,y,z_0;t)\ \epsilon^{-1}\ d\epsilon\, dy
 \label{tcb-append-eq-int}
\end{equation}
and $\epsilon_0=\left(m\rs{e}c^2\right)^2/E\rs{\gamma}$, $\sigma_{\gamma\gamma}$ is the cross-section of the interaction \citep{1983Ap.....19..187A}, $y$ is the coordinate along the line of sight, $(x_0,y_0,z_0)$ are the coordinates of the cell where the \g-photons with energy $E\rs{\gamma}$ were emitted and $(x_0,y_1,z_0)$ are the coordinates of the nova shock closer to the observer. 

Since the integration is performed along the line of sight and accounts for internal structures, the absorption may depend on the orientation of the system with respect to the observer in the presence of asymmetry or anisotropy.

\end{appendix}  

\end{document}